\documentclass[iop,apj]{emulateapj}

\usepackage{multirow}
\usepackage{txfonts}
\usepackage{graphicx}
\def\ltsima{$\; \buildrel < \over \sim \;$}
\def\simlt{\lower.5ex\hbox{\ltsima}}
\def\gtsima{$\; \buildrel > \over \sim \;$}
\def\simgt{\lower.5ex\hbox{\gtsima}}

\def\cm2{{cm$^{-2}$}}

\def\lum{{$L_{0.5-10}$}}

\def\p1{{Paper I}}

\def\xmm{{\em XMM--Newton}}
\def\chandra~{{\em Chandra}}

\def\chandra{{\em Chandra}}

\def\xmm{{\em XMM--Newton}}

\def\f14{{10$^{-14}$}}
\def\f13{{10$^{-13}$}}
\def\f12{{10$^{-12}$}}
\def\f11{{10$^{-11}$}}
\def\e22{{10$^{22}$}}

\def\3c{{3C 234}}
\def\l58{{$L_{5.8 \mu m}$}}
\def\sig{$\sigma^{2}_{rms}$}
\def\sigv{$\sigma^{2}_{rms,V>1.3}$}

\slugcomment{Accepted to the Astrophysical Journal on December 6, 2013}

\shorttitle{Variability in XMM-COSMOS}
\shortauthors{Lanzuisi et al.}

\begin{document}

\title{AGN X-ray variability in the XMM-COSMOS survey}

\author{G. Lanzuisi\altaffilmark{1,2} ; G. Ponti\altaffilmark{1}; M. Salvato\altaffilmark{1,3}; G. Hasinger\altaffilmark{4}; 
N. Cappelluti\altaffilmark{5}; A. Bongiorno\altaffilmark{6}; M. Brusa\altaffilmark{1,7}; E. Lusso\altaffilmark{8};
P. K. Nandra\altaffilmark{1}; A. Merloni\altaffilmark{1,3}; J. Silverman\altaffilmark{9}; J. Trump\altaffilmark{10}; C. Vignali\altaffilmark{7}; A. Comastri\altaffilmark{5};
R. Gilli\altaffilmark{5}; M. Schramm\altaffilmark{9}; C. Steinhardt\altaffilmark{8}; 
D. Sanders\altaffilmark{4}; J. Kartaltepe\altaffilmark{11}; D. Rosario\altaffilmark{1}; B. Trakhtenbrot\altaffilmark{12}}

\altaffiltext{1}{Max-Planck-Institut f\"ur extraterrestrische Physik,  Giessenbachstrasse, 85748 Garching, Germany}
\altaffiltext{2}{TUM Fakult\"at f\"ur Physik, James-Franck-Strasse, 185748 Garching, Germany}
\altaffiltext{3}{Excellence Cluster, Boltzmann Strasse 2, 85748 Garching, Germany}
\altaffiltext{4}{Institute for Astronomy, 2680 Woodlawn Drive, Honolulu, HI 96822-1839, USA}
\altaffiltext{5}{INAF - Osservatorio Astronomico di Bologna, via Ranzani 1, 40127 Bologna, Italy}
\altaffiltext{6}{Istituto Nazionale di Astrofisica - Osservatorio Astronomico di Roma Via Frascati 33, 00040, Monte Porzio Catone, Italy}
\altaffiltext{7}{Dipartimento di Fisica e Astronomia, Universit\'a di Bologna, viale Berti Pichat 6/2, 40127 Bologna, Italy}
\altaffiltext{8}{Max Planck Institut f\"ur Astronomie, K\"onigstuhl 17 D-69117 Heidelberg, Germany}
\altaffiltext{9}{Kavli Institute for the Physics and Mathematics of the Universe (Kavli IPMU) 5-1-5 Kashiwanoha Kashiwa, 277-8583, Japan}
\altaffiltext{10}{University of California Observatories/Lick Observatory and Department of Astronomy and Astrophysics, University of California}
\altaffiltext{11}{National Optical Astronomy Observatory, 950 North Cherry Avenue, Tucson, AZ 85719, USA}
\altaffiltext{11}{Wise Observatory Tel Aviv University 69978 Tel Aviv, Israel POB: 39040}


\begin{abstract}

We took advantage of the observations carried out by XMM in the COSMOS field during 3.5 years, 
to study the long term variability of a large sample of AGN (638 sources),
in a wide range of redshift ($0.1<z<3.5$) and X-ray luminosity ($10^{41}<$\lum$<10^{45.5}$).
Both a simple statistical method to asses the significance of variability, and the 
Normalized Excess Variance (\sig) parameter, where used to obtain a quantitative measurement of the variability.
Variability is found to be prevalent in most AGN, whenever we have good
statistic to measure it, and no significant differences between type-1 and type-2 AGN were found.
A flat (slope $-0.23\pm0.03$) anti-correlation between \sig\ and X-ray luminosity is found,
when significantly variable sources are considered all together.
When divided in three redshift bins, the anti-correlation becomes stronger 
and evolving with z, with higher redshift AGN being more variable.
We prove however that this effect is due to the pre-selection of variable sources:
considering all the sources with available \sig\ measurement, the evolution in redshift disappears.
For the first time we were also able to study the long term X-ray variability as a function of $M_{\rm BH}$ and Eddington ratio, 
for a large sample of AGN spanning a wide range of redshift. 
An anti-correlation between \sig\ and $M_{\rm BH}$ is found, with the same slope of the anti-correlation
between \sig\ and X-ray luminosity, suggesting that the latter can be a byproduct of the former one.
No clear correlation is found between \sig\ and the Eddington ratio in our sample.
Finally, no correlation is found between the X-ray \sig\ and the optical variability.
\end{abstract}

\section{Introduction}

Variability, on timescales from minutes to years,
is one of the defining characteristics of AGN/BH accretion.
Indeed, the rapid variability of quasar was one of the arguments for the presence of
a compact central engine powering these sources (Rees 1984).
Variability has long been used as an AGN selection technique in the optical (van den Bergh et al. 1973;
Bonoli et al. 1979; Giallongo et al. 1991; Trevese et al. 1994; Vanden Berk et al. 2004; de Vries
et al. 2005).
More recently optical variability has been used to select AGN (typically Low Luminosity AGN) 
in X-ray and multi-wavelength surveys, to complement other selection techniques
(Trevese et al. 2008; Villforth et al. 2010; Sarajedini et al. 2011, Young et al. 2012).

In the X-rays, first results from EXOSAT and RXTE showed that the variability amplitude is anti-correlated with X-ray luminosity,
and that the power spectral density (PSD) can be modeled with a power law with slope steeper than 1,
i.e. the variability decreases with increasing frequency
(Barr \& Mushotzky 1986; Lawrence \& Papadakis 1993, Green, McHardy \& Lehto 1993; Nandra et al. 1997; Markowitz \& Edelson 2004).

The combination of these results with short term, high quality light curves with \xmm, for a handful of AGN, 
allowed the study of the PSD in a larger range of frequencies.
As previously suggested (Papadakis \& McHardy 1995) the steep PSD flattens below a break frequency $\nu_{b}$
(Edelson \& Nandra 1999; Uttley et al. 2002; Markowitz \& Edelson 2004).
A similar behavior is observed in galactic BH binaries (BHBs, e.g. Axelsson et al. 2005; Gierlinski et al. 2008).
This suggests that similar processes are in place for these two classes of sources,
with the difference in time scales related to the different BH mass ($M_{\rm BH}$) ranges involved.
Thus, the variability-luminosity relation could be a consequence of an intrinsic
variability-BH mass relation (Hayashida et al. 1998; Czerny et al. 2001; Papadakis 2004).

McHardy et al. (2006) demonstrated that, as far as variability is concerned,
SMBH are scaled versions of BHB and that, over 8 orders of magnitude in $M_{\rm BH}$ and 6 in frequency, 
the  $\nu_{b}$  is inversely correlated with the BH mass, once it is corrected for different accretion rates.
Koerding et al. (2007) proposed a direct correlation between  $\nu_{b}$, $M_{\rm BH}$
and accretion rate ($\dot{M}$), where $\nu_{b}$ scales linearly with both $\dot{M}$ and $M_{\rm BH}^{-1}$ 
 (but see also Gonz{\'a}lez-Mart{\'{\i}}n \& Vaughan 2012).

For variability studies, the PSD is a powerful tool
but long, uninterrupted light curves for AGN are difficult to achieve and are available only for a handful of local sources.
The normalized excess variance \sig\ has been used as a more convenient tool for sparse sampling of light curves in large AGN samples.
Nandra et al. (1997), Turner et al. (1999) and George et al. (2000) applied this technique to ASCA data, to both quasars and Seyferts, 
finding an anti-correlation of \sig\ with X-ray luminosity and $M_{\rm BH}$, and a correlation with X-ray spectral index.
O'Neill et al. (2005) used ASCA light curves of 46 local AGN to study the correlation with $M_{\rm BH}$ (and $L_{2-10}$).
Zhou et al. (2010) found, on a sample of local AGN with available reverberation mapping BH masses, 
a tight correlation between \sig\ and $M_{\rm BH}$, proposing it as a method to infer the BH mass from \sig.
Using a sample of 161 local (mostly at $z<0.2$) AGN 
(every X-ray unabsorbed radio quiet AGN observed by XMM in pointed observations), 
Ponti et al. (2012, hereafter P12) found that, to first approximation, all local AGN have the same variability properties once 
rescaled for $M_{\rm BH}$ (and $\dot{M}$). The authors measured a tight correlation between \sig\ and $M_{\rm BH}$, 
with the scatter becoming smaller (only a factor of 2-3, comparable to the one induced by the $M_{\rm BH}$ uncertainties) 
when the subsample of reverberation mapped AGN was considered  (see also Kelly et al. 2011). This suggests that X-ray variability $M_{\rm BH}$ measurements 
are more accurate than the ones based on single epoch optical spectra. 
A highly significant correlation between \sig\ and 2-10 keV spectral index was also observed. 

All these studies are based on local AGN and short time scales light curves (typically from minutes to days),
i.e. in a range of frequencies typically above  $\nu_{b}$. 
 Given that \sig\ roughly measures the integral of the PSD, over  the sampled time  scales, 
the shift of $\nu_{b}$ with the BH mass it thought to produce
the correlation between the observed variability and $M_{\rm BH}$.
On longer timescales, therefore at $\nu<<\nu_{b}$, the PSD has been measured only for a few AGN. 
In particular, the normalization of the flat portion of the PSD is not well know. 
It is generally assumed that this normalization is the same for all AGN, 
regardless of the BH mass and luminosity (e.g Papadakis 2004),
as recently observed on very long time scales RXTE light curves of a small sample of local AGN (Zhang 2011).
The  $\nu_{b}$ for a BH mass of $10^9$ M$_{\odot}$ is of the order of 1 year, so very long observations/samplings are
required to investigate this frequency range.

This range of frequencies is becoming accessible for large sample of sources 
in deep X-ray surveys, where the large total exposure is achieved  through repeated observations over years.
They are able to detect variability in moderate-luminosity  or high redshift AGN
(e.g. Paolillo et al. 2004 on CDF-S, Papadakis et al. 2008 on Lockman Hole data).
Young et al. (2012) utilized X-ray variability to select elusive AGN (finding 20 new AGN candidates), using
the 4 Msec of \chandra\ exposure in the CDF-S, taken over 10.8 years. 

In this paper we use the repeated observations performed with \xmm\ to cover the 2 deg$^2$ field of COSMOS (Scoville et al. 2007),
with 55 pointings from Dec. 2003 to May 2007 and a total of 1.5 Msec of exposure (Hasinger et al. 2007),
to study the long term variability of the entire catalog of X-ray detected sources (Cappelluti et al. 2009).
The catalog has an almost 100\% completeness in optical/IR identification (Brusa et al. 2010).
The availability of spectroscopic redshifts or reliable photometric redshifts 
allows us to study the X-ray variability as a function of spectral type, luminosity  and redshift. 
Most ($\sim800$) of our sources have a L$_{Bol}$ estimate from SED fitting (Lusso et al. 2011; 2012),
and a BH Mass estimate, either from optical broad line for type-1 
(Trump et al. 2009; Merloni et al. 2010; Matsuoka et al. 2013), or from scaling relations for type-2 (Lusso et al. 2011),
allowing us to study the dependency of the variability also as a function of BH mass and accretion rate.

The paper is organized as follows: 
section 2 summarizes the properties of the data set and the light curves extraction;
section 3 presents the variability estimator V, while section 4 introduces the Normalized Excess Variance (\sig);
in section 5 the distributions of \sig\ as a function of \lum, z, $M_{BH}$, Eddington ratio and optical variability are discussed;
section 6 summarizes the results.
A standard $\Lambda$ cold dark matter cosmology with $H_0=70$ km s$^{-1}$ Mpc$^{-1}$, $\Omega_\Lambda=0.73$ and $\Omega_M=0.27$ is assumed throughout the paper.
Errors are given at 1 $\sigma$ confidence level for one interesting parameter, unless otherwise specified.

\begin{figure*}[t]
\begin{center}
\includegraphics[width=8cm,height=8cm]{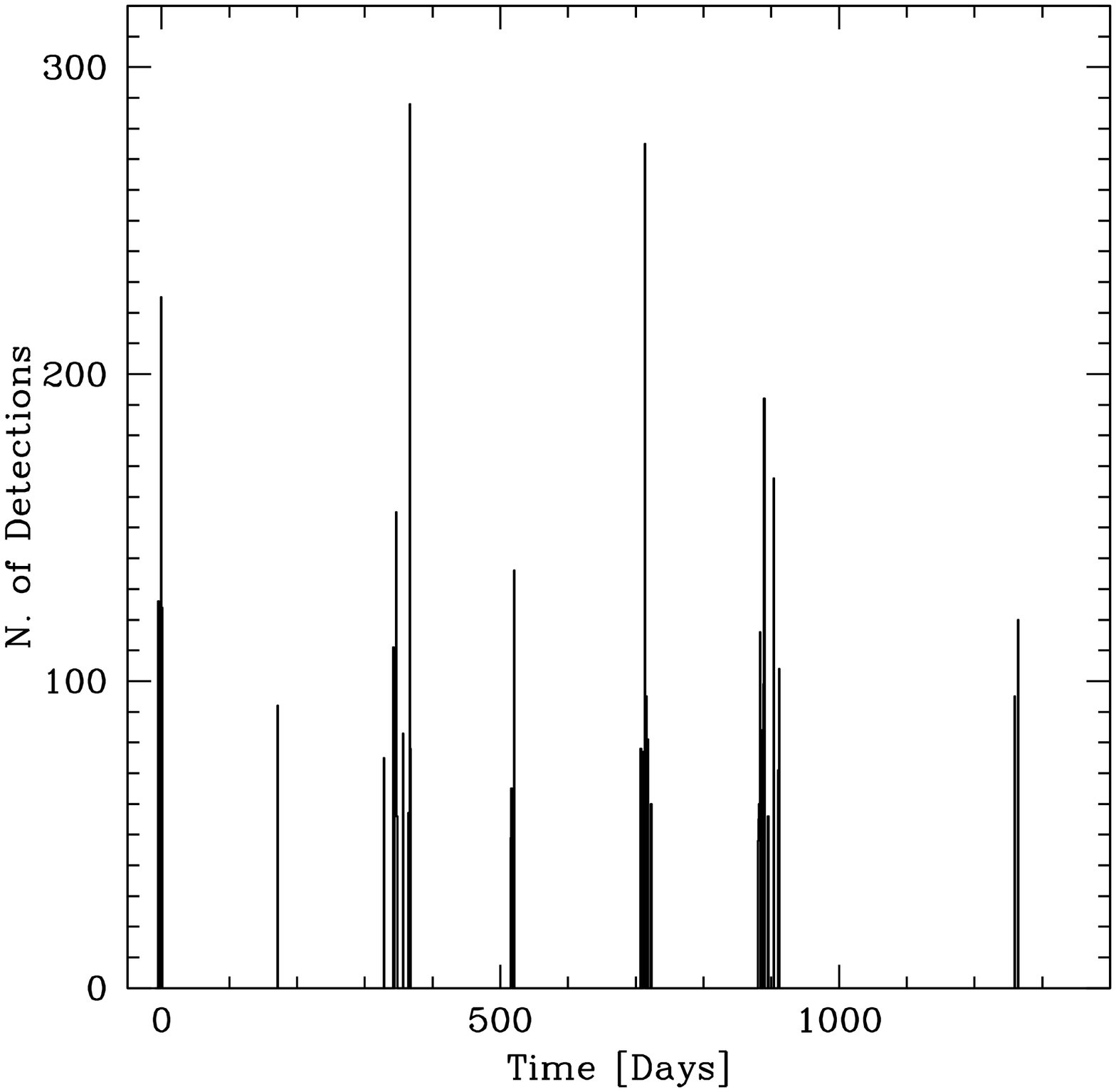}\hspace{0.5cm}
\includegraphics[width=8cm,height=8cm]{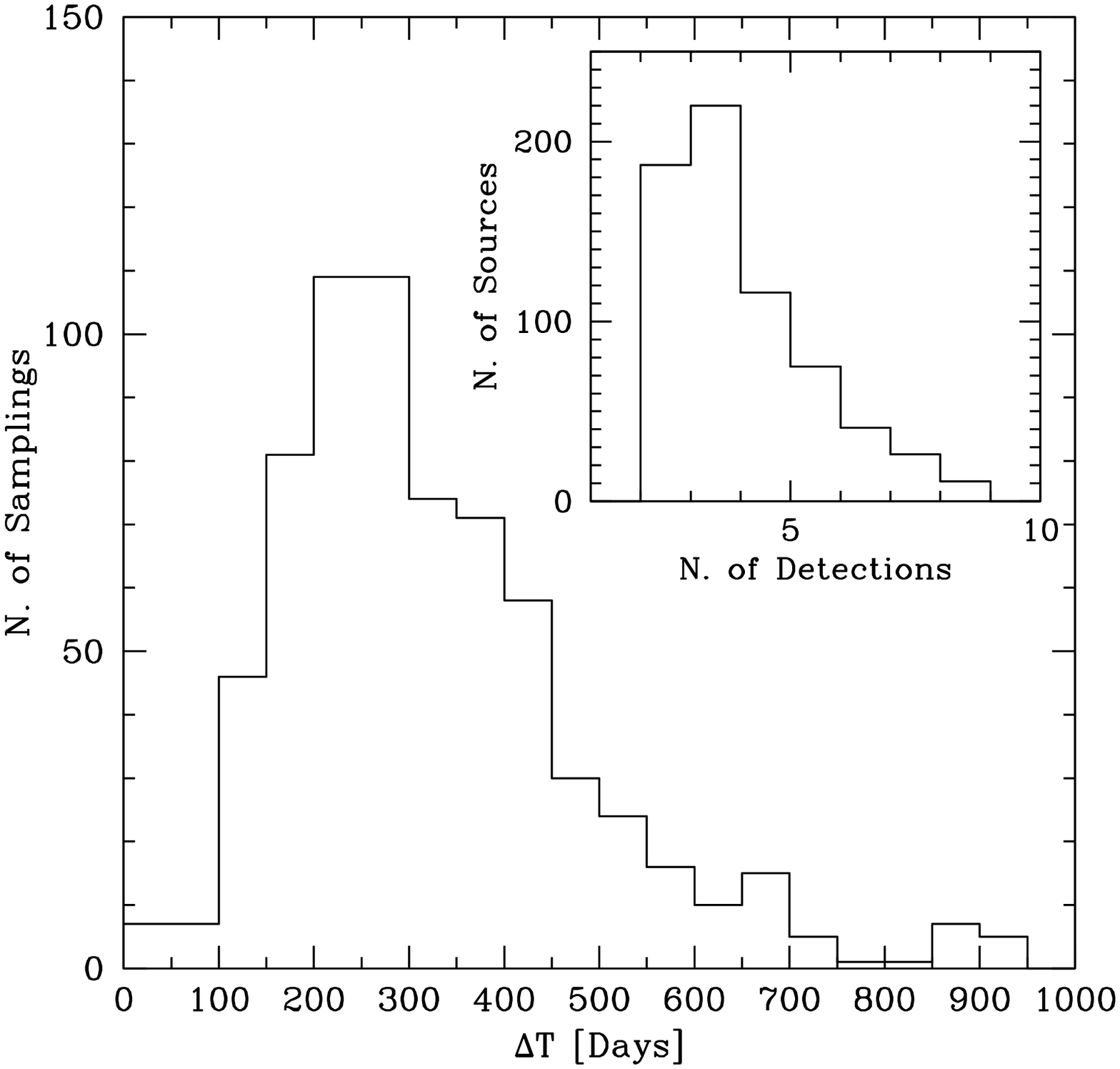}
\caption{{\it Left panel:} Distribution of single source detections,
during the 3.5 years observing campaign (between 2003-12-06 and 2007-05-018). 
{\it Right panel:} Distribution of the total lightcurve length, rescaled for the redshift of the source, 
for all the sources. The inset show the distribution in number of detections
for sources in our sample.
}
\end{center}
\label{times}
\end{figure*}

\section{Data set}

\subsection{The COSMOS survey}

The Cosmic Evolution Survey (COSMOS) is a deep and wide
extragalactic survey designed to probe the formation and evolution of galaxies as a function of cosmic time 
and large scale structural environment. 
The survey covers a 2~deg$^2$ equatorial (10$^h$, +02$^{\circ}$) field with imaging by most of the major space-based telescopes 
( {\em Hubble} , {\em Spitzer}, GALEX, \xmm, \chandra) 
and a number of large ground based telescopes ({\em Subaru}, VLA, ESO-VLT, UKIRT, NOAO, CFHT, and others).
Large dedicated ground-based spectroscopy
programs in the optical with Magellan/IMACS (Trump et al. 2007), VLT/VIMOS
(Lilly et al. 2009), {\em Subaru}-FMOS and DEIMOS-{\em Keck} have been completed or are well underway.

This wealth of data has resulted in an 35-band photometric catalog of
$\sim$10$^6$ objects (Capak et al. 2007) resulting in
photo-z's for the galaxy population accurate to $\Delta$z/(1+z)$<$1\%
(Ilbert et al. 2009) and to $\Delta$z/(1+z)$\sim$1.5\% for the AGN population
(Salvato et al. 2009, 2011).

\subsection{XMM-COSMOS}

The \xmm\ wide-field survey in the COSMOS field
(hereinafter XMM-COSMOS; Hasinger et al. 2007) is a crucial
component of the multi-wavelength coverage of the COSMOS field.
The $\sim2$ deg$^2$ area of the HST/ACS program
has been surveyed with \xmm\ for a total of $\sim1.55$ Ms
during AO3, AO4, and AO6 cycles of XMM observations,
providing an unprecedented large sample of point-like X-ray sources ($\sim1800$).

The XMM-COSMOS project is described in Hasinger et al. (2007, Paper I), while the X-ray point source catalog
and counts from the complete XMM-COSMOS survey are presented in Cappelluti et al. (2009, Paper II). 
Brusa et al. (2010, Paper III) presents the optical identifications of the X-ray point sources
in the XMM-COSMOS survey and the multi-wavelength properties of this large sample of X-ray selected AGNs.
The catalog used in this work includes 1797 point-like sources
detected in the 0.5-2 (1545 sources), 2-8 (1078 sources) 
and 4.5-10 (246 sources) keV bands. 
The nominal limiting fluxes are $\sim5\times10^{-16}$ , $\sim3\times10^{-15}$ , and $\sim7\times10^{-15}$ erg cm$^{-2}$ s$^{-1}$ ,
respectively. The adopted likelihood threshold corresponds
to a probability $1\times10^{-10}$ that a catalog source is a spurious 
background fluctuation. In the present analysis, we used the source list
created from the 53 XMM-COSMOS fields.

The XMM-COSMOS catalog has almost 100\% redshift completeness (Paper III). 
884 (out of 1797) sources have a spectroscopic
redshift, 748 sources have a photometric redshift, 97 are classified as stars and 68 remain unclassified. 
The sources with spectroscopic redshift are divided almost equally between broad line 
(FWHM$>2000$ km/s, BL) and non-BL AGN, 
plus a small fraction of non AGN sources, i.e. passive galaxies at low redshift or stars.
The sources for which only photometric redshift is available, have been 
classified on the basis of the best SED fitting template.
The agreement between SED classification and spectral classification is good 
(Lusso et al 2010, Salvato et al. 2009, Brusa et al. 2010).
The final classification breakdown is: 611 type-1, 941 type-2, 80 Galaxies, 97 Stars, 68 Unclassified.

We underline that, in the following analysis, with type-1 we refer to sources that either have a spectroscopic redshift, showing at least one broad
emission line, or have as best fit SED templates an unobscured quasars template
with different degrees of contamination by star-forming galaxy templates (from 10 to 90\% of the Optical-NIR flux).
With type-2 we refer to sources with optical spectra of narrow lines AGN or passive/star-forming galaxies,
or best fit SED templates of obscured AGN, with different degrees of contamination by
passive or star-forming galaxy templates.
We exclude, from the following analysis, sources classified as starts or unclassified.

\subsection{Light curve extraction}

Because of the observations pattern, sources in XMM-COSMOS can be observed in a minimum of 1 to a maximum of
11 pointing. We have the possibility of following up
every source for a period of up to 3.5 years. 
By using the XMM-SAS tool {\it emldetect} we selected all those sources
with a detection significance $det_{ml}>10$ (corresponding to a probability that a Poissonian fluctuation of the
background is detected as a spurious source of $P=10^{-10}$),
in each pointing. We then produced light curves
of all those sources observed in more than 1 pointing. As a result
we produced light curves for 995 of the 1797 sources in the full band catalog. 
The number of detections spans from 2 to 9, depending 
on the position of the source and on its flux.
Fig. 1 (left panel) shows the distribution of all the detections of  sources in the total catalog,
during the 3.5 years observing campaign (between 2003-12-06 and 2007-05-18). 
We have a total of 3849 single detections, distributed among the observations depending on the exposure 
time of each pointing.
In order to study variability in these sources we limited the following analysis to
sources with $\geq3$ or more detections.

To have an estimate of the typical rest frame time scales sampled by the available 3.5 years observing campaign,
we computed, for each source, the time intervals between the first and the last detections, 
rescaled for the redshift of the source, for time
dilation, i.e. $\Delta t_{rest}=\Delta t/(1+z)$. 
Fig. 1 (right panel) shows the distribution of the time intervals for all the  638 sources
with $\geq3$ detections.
The inset shows the distribution of the number of detections for each source in the sample:
190 sources have only 3 detections in total, and in the last bin we have 10 sources with 9 detections, which is
the maximum number of samplings available.
The plot shows that, given the distribution of the observations, 
and the redshift distribution of the sources,
 the vast majority of the light curves have a total length in the range 100-500 days (rest-frame), 
i.e. the low frequency limit of the lightcurve is 
typically at frequencies lower than $\nu_b$ for most of our sources.
As reference, we recall that the typical break frequency in the PSD of an AGN with a  $10^8$ M$_\odot$ BH
is $1/\nu_b\sim30$ days.
The final breakdown of sources with light curves is 340 type-1, 291 type-2, 7 Galaxies.

The light curves are measured in the fixed 0.5-10 keV observed band, implying that 
we are measuring variability at increasingly high intrinsic energies, up to 2-40 keV at $z=3$.
Even if variability can be in principle highly energy dependent, given the different spectral components,
(with different origins) which dominate different bands,
observationally, it is known to be well correlated between the 0.3-0.7, 0.7-2 and 2-10 keV bands (e.g. P12).
In soft X-rays the effect of observing variability in different bands is small:
typically $\sim10\%$ from 0.5 to 10 keV for local Seyferts (Vaughan et al. 2004, Gallo et al. 2007). 
In hard X-rays (above 10 keV) there are less data available. However, Caballero-Garcia et al. (2012) found that 
the variability amplitude above 10 keV is similar to the one below 10 keV, for a sample of 5 bright AGN in
the Swift-BAT 58 month catalog (see also Beckmann et al. 2008; Soldi et al. 2011).
Soldi et al. (2013 submitted) reported, for a sample of 110 radio quiet QSO from the same Swift-BAT 58 month catalog,
that the variability is very well correlated even between the 14-24 and 35-100 keV bands (with again 10\% difference).
Furthermore, all the variability estimators are strongly dependent on the available number of counts (see Sec. 3), 
hence we decided to exploit all the X-ray counts available using the full \xmm\ observing band for all sources,
instead of restricting ourselves to a much narrower fixed res-frame band. 
It is true, however, that we do not know if the variability is energy dependent in AGN at high redshift, 
contrary to what is observed in the local Universe. To answer this question however, one needs a large sample
of sources, spanning a large redshift range, and not so severely affected by low signal to noise ratio.
The almost  uniform 200 ks \chandra\ coverage on the entire 2 deg$^2$ COSMOS field, that will be available by the end of 2013,
will be the perfect data set for such a study.


\begin{figure*}
\begin{center}
\includegraphics[width=8cm,height=8cm]{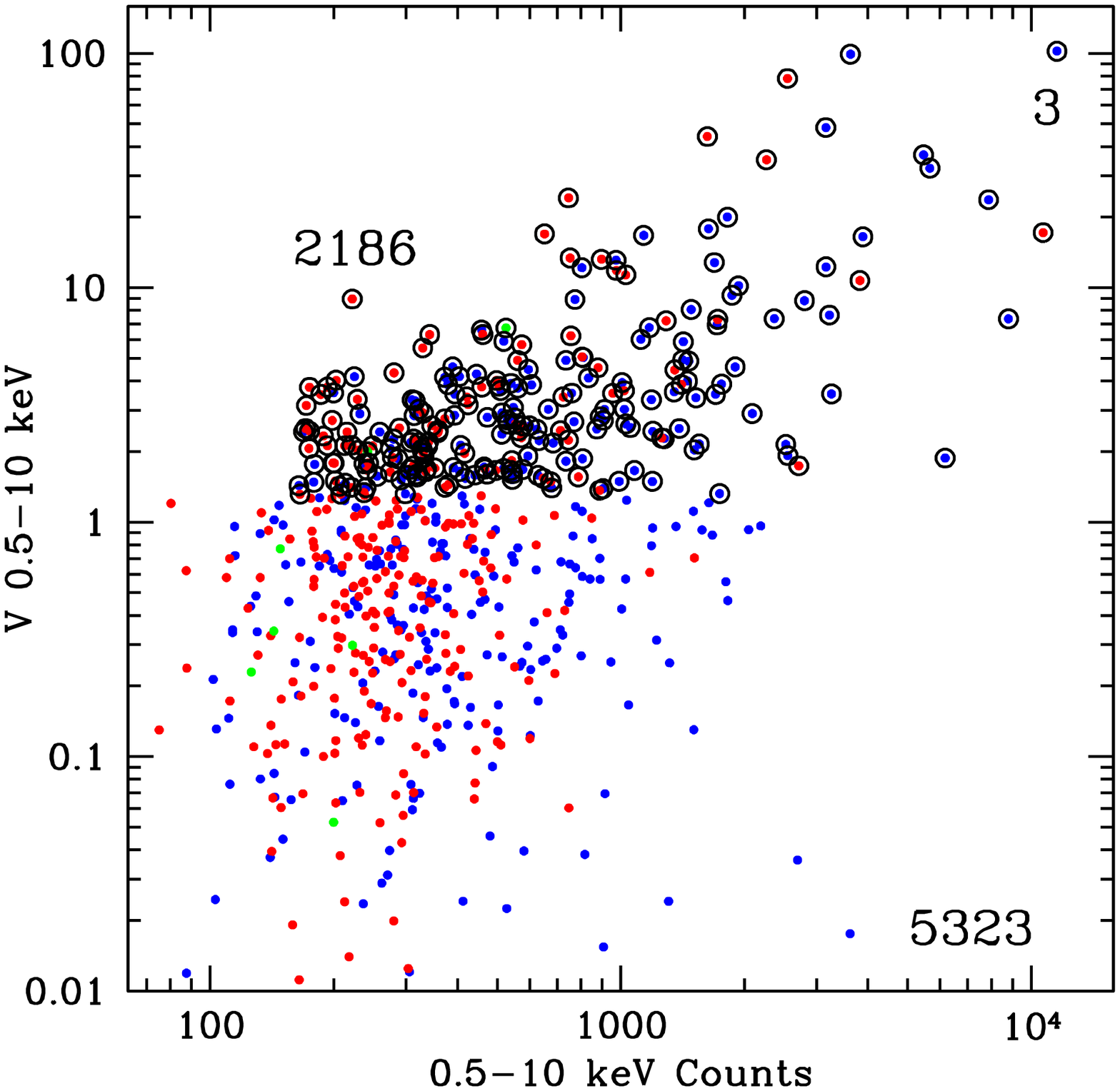}\hspace{0.5cm}
\includegraphics[width=8cm,height=8cm]{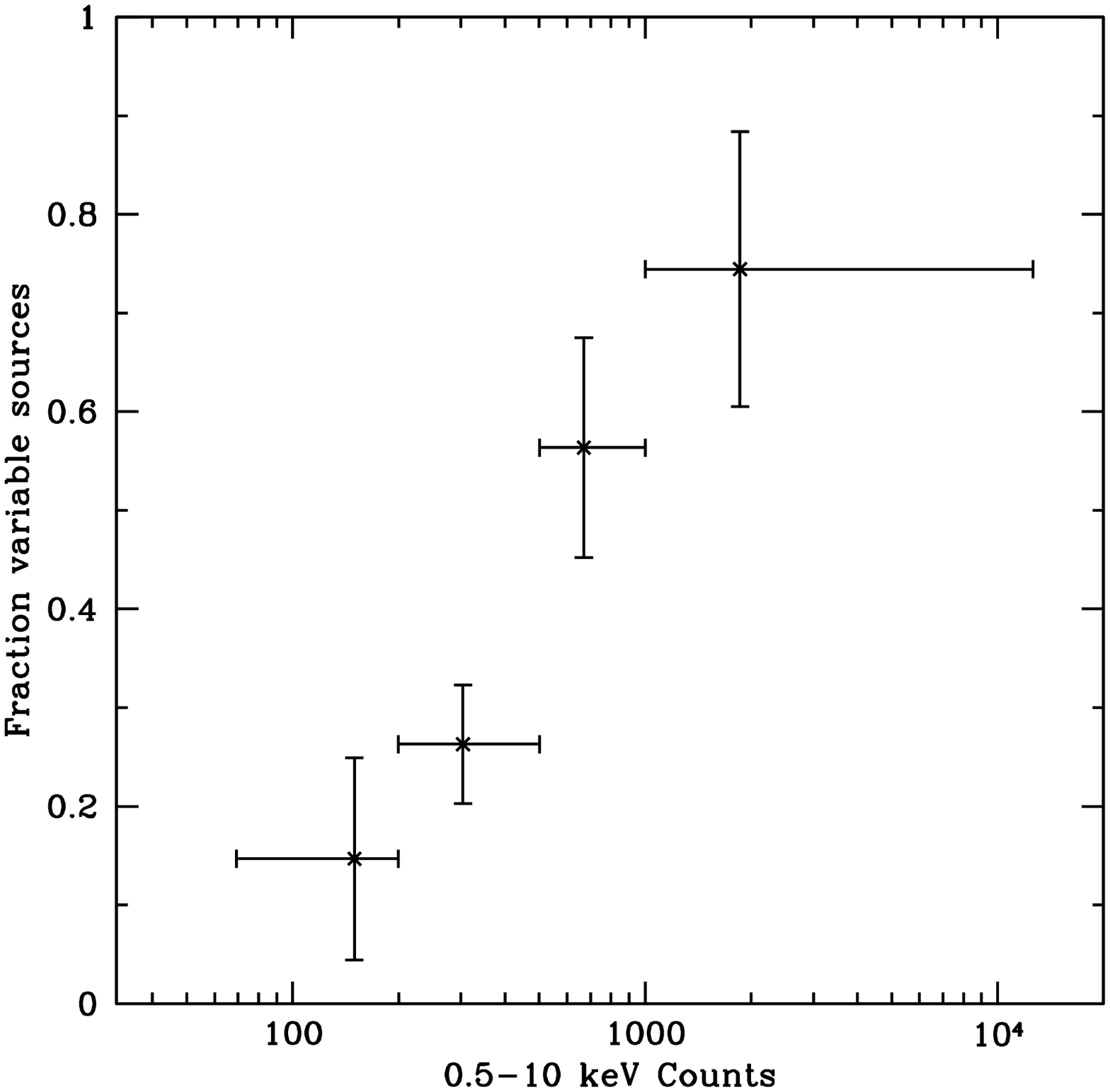}
\caption{{\it Left panel:} Distribution of V parameter as a function of 0.5-10 keV net counts. 
Blue are type-1, red type-2 and green galaxies. Sources with $V > 1.3$ are labeled with black circles. 
This corresponds to a source being variable at 95\% confidence level.
Three peculiar sources are labeled with their XID number. Their X-ray and optical light curves are shown in Fig. 3.
{\it Right panel:} Fraction of variable sources ($V > 1.3$) as a function of number of counts, in bins of counts.}
\end{center}
\label{vcounts}
\end{figure*}

\begin{figure*}
\begin{center}
\includegraphics[width=5.5cm,height=7cm]{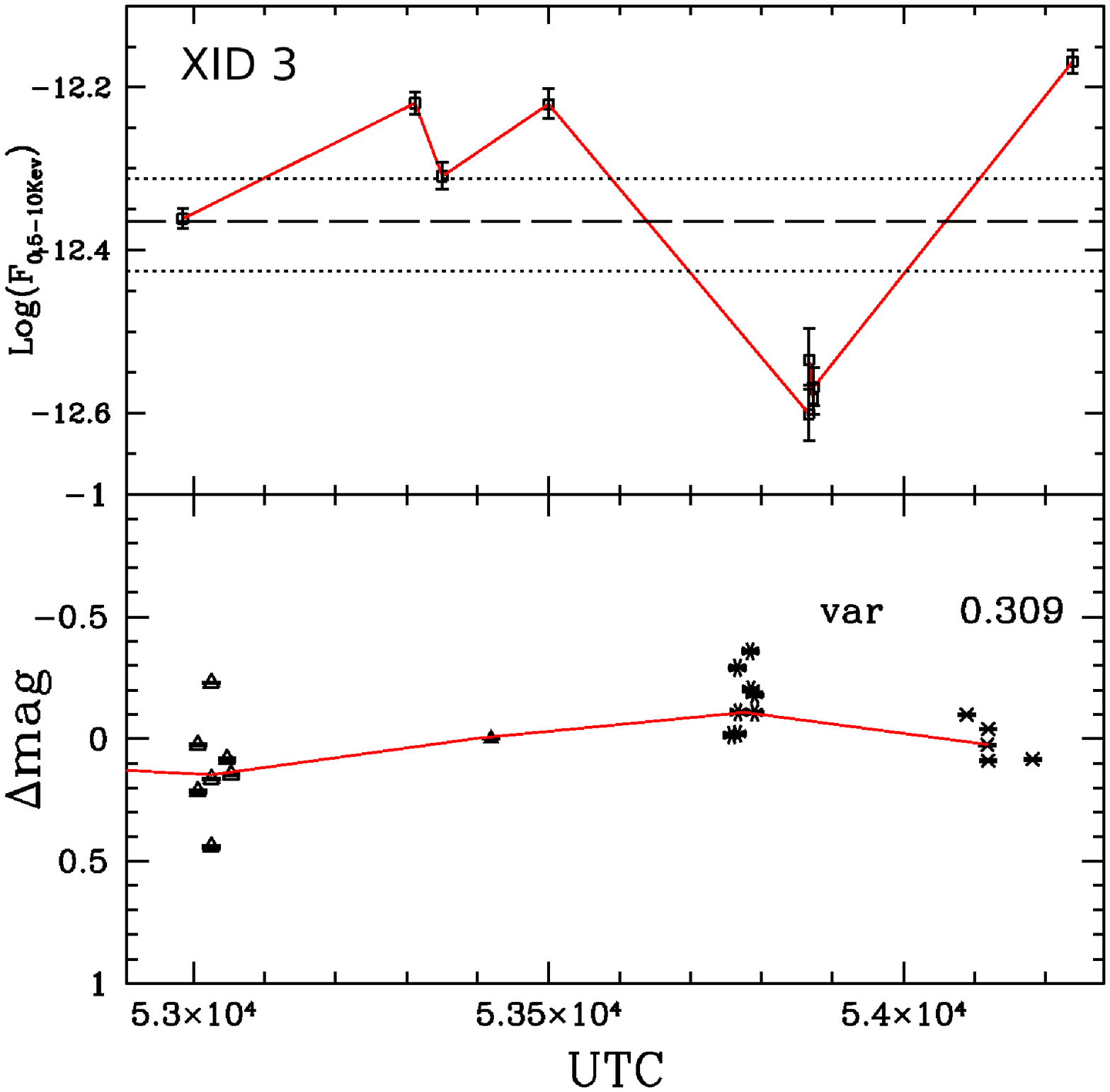}\hspace{0.5cm}\includegraphics[width=5.5cm,height=7cm]{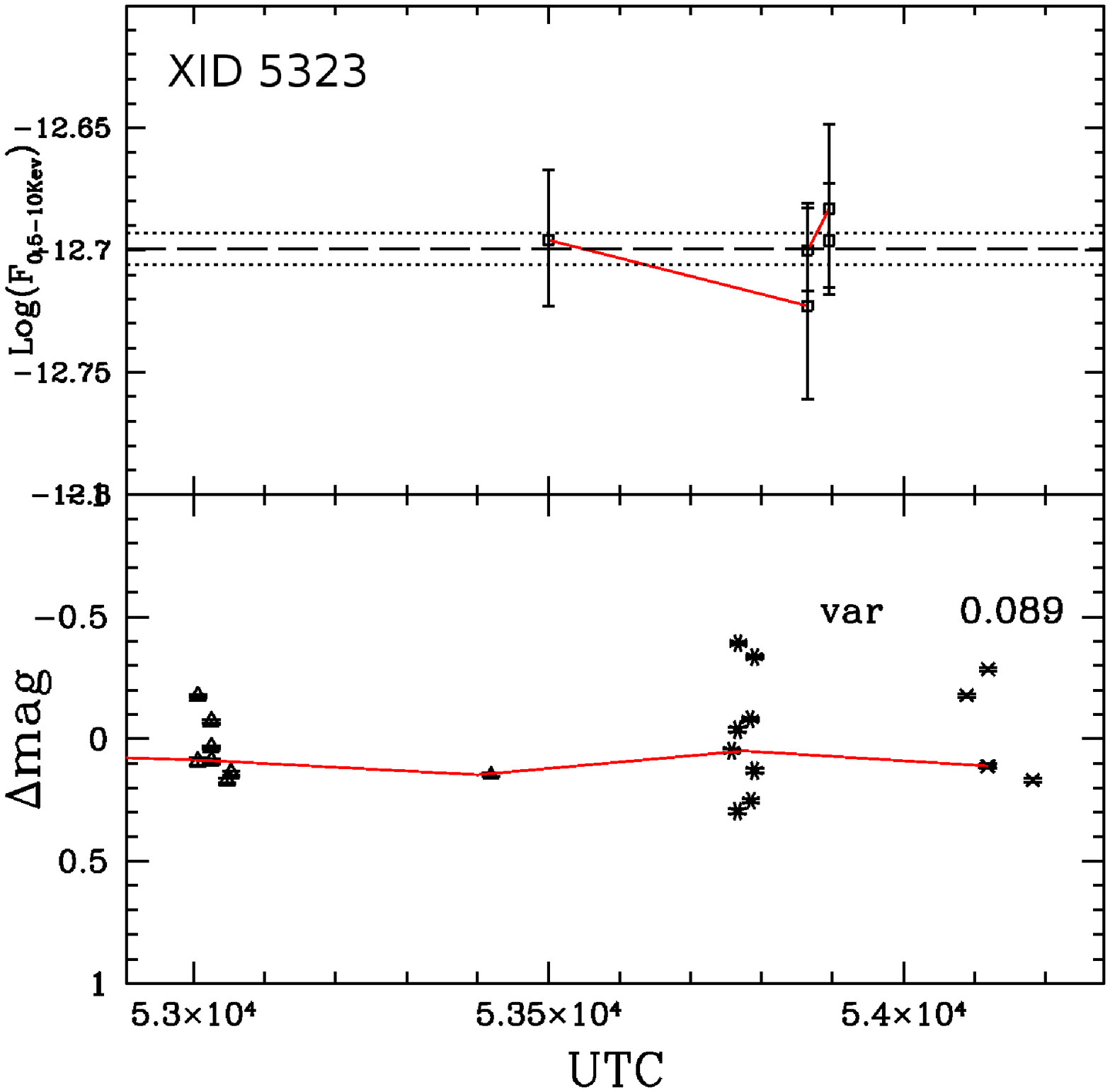}\hspace{0.5cm}\includegraphics[width=5.5cm,height=7cm]{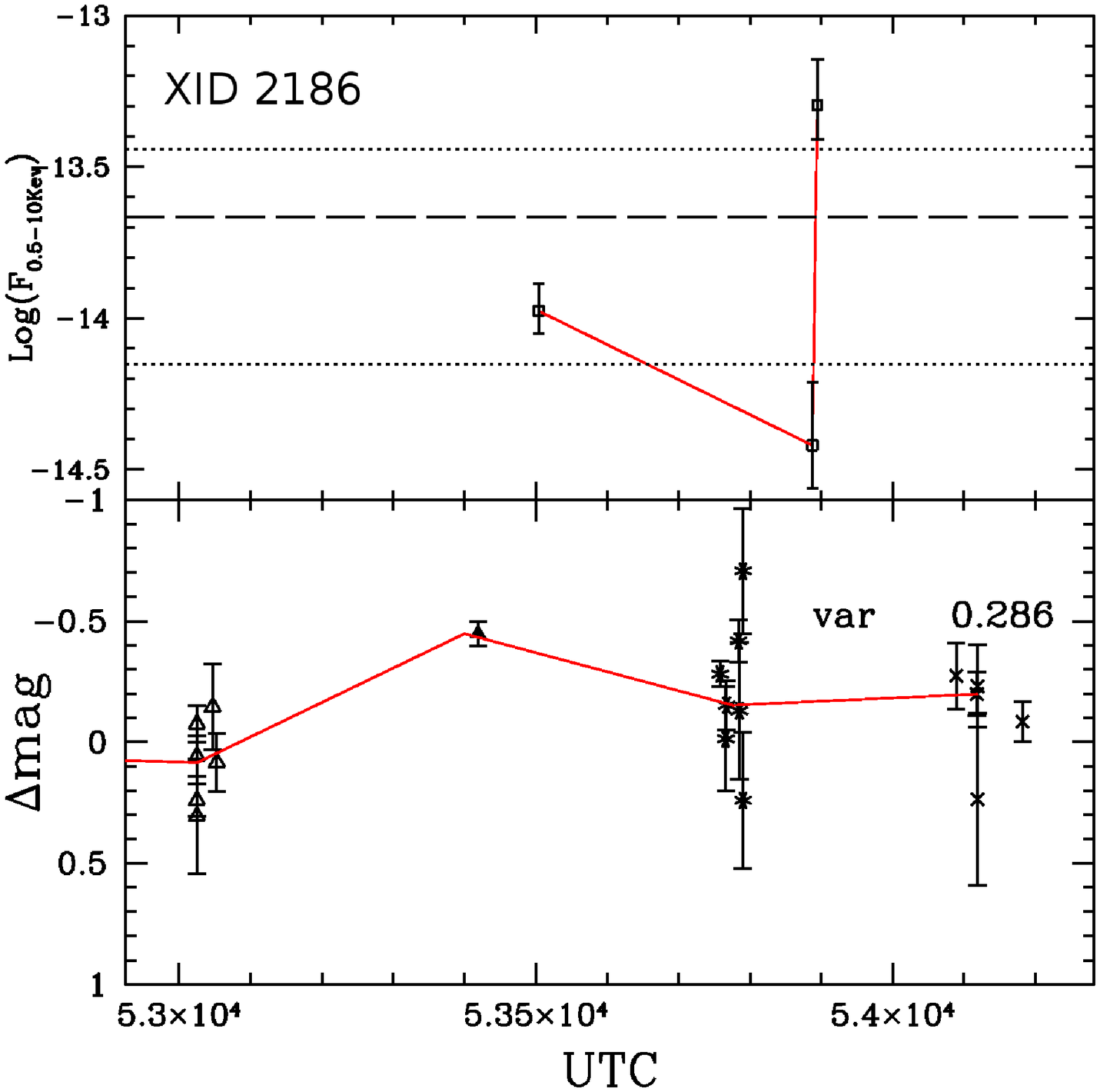}
\caption{X-ray (top) and Optical (bottom) time variability for the sources XID 3, 5323, and 2186, respectively.
{\it Top panels:} The dashed line shows the weighted mean of the 0.5-10 keV flux. 
Dotted lines show the standard error on the mean. The continuous line connects data points.
{\it Bottom panels:} The sources was observed in 4 epochs, each epoch marked with a different symbol. 
The four groups are, from left to right: Subaru broad-band images (Subaru$_{BB}$), CFHT$ {\it g}$ band, 
the first set of intermediate bands Subaru images (Subaru$_{IB1}$), and the second epoch of intermediate bands 
Subaru images (Subaru$_{IB2}$). 
The red line connects the median values of the deviation from a running Gaussian filter, for each group of observations. 
The points in each group refer to different wavelengths, therefore the difference within points in each group are due to 
the SED shape, and not variability.
More details on the optical variability in Salvato et al 2009.
}
\end{center}
\label{lc}
\end{figure*}

\begin{figure*}[t]
\begin{center}
\includegraphics[width=8cm,height=8cm]{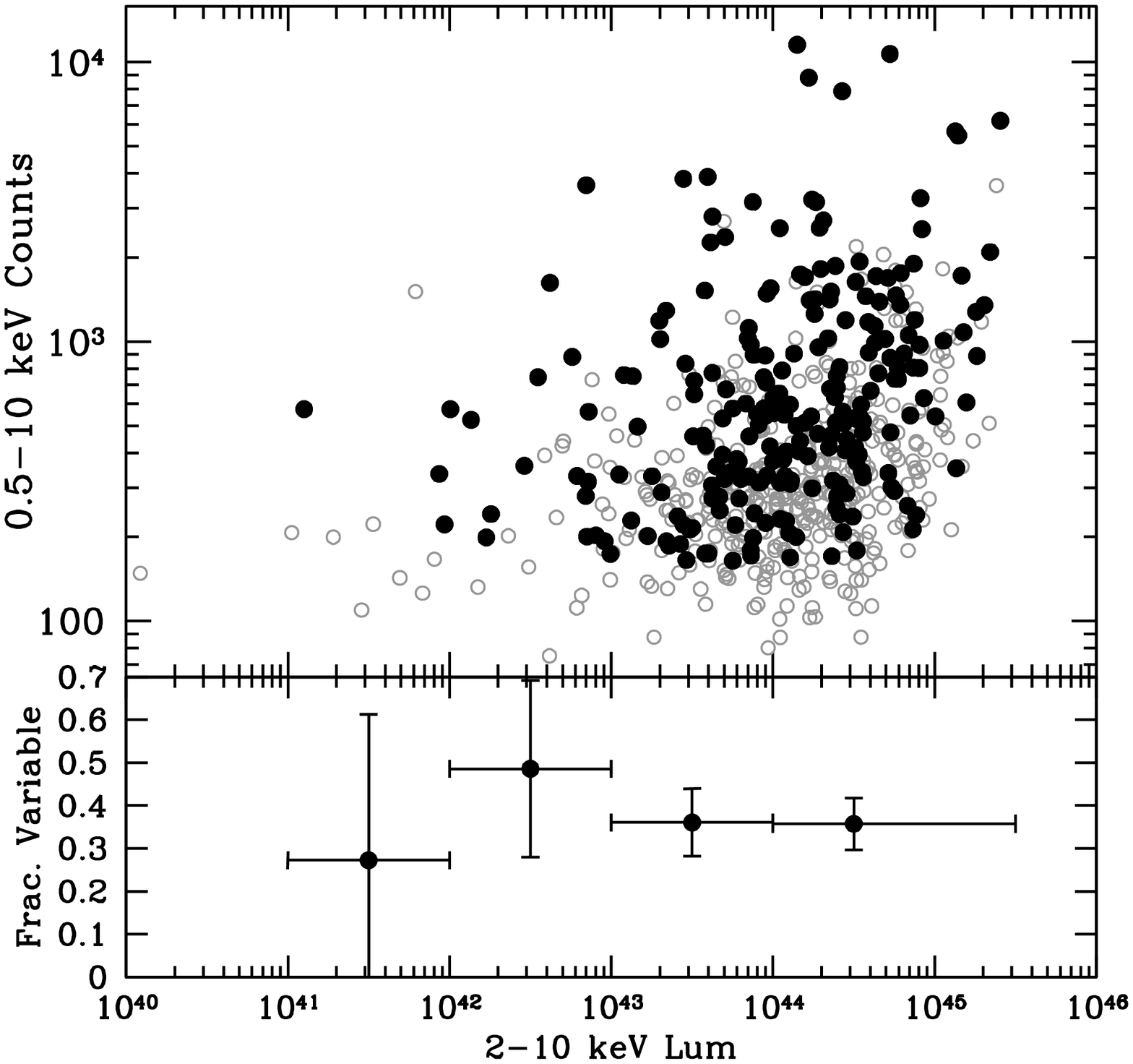}\hspace{0.5cm}
\includegraphics[width=8cm,height=8cm]{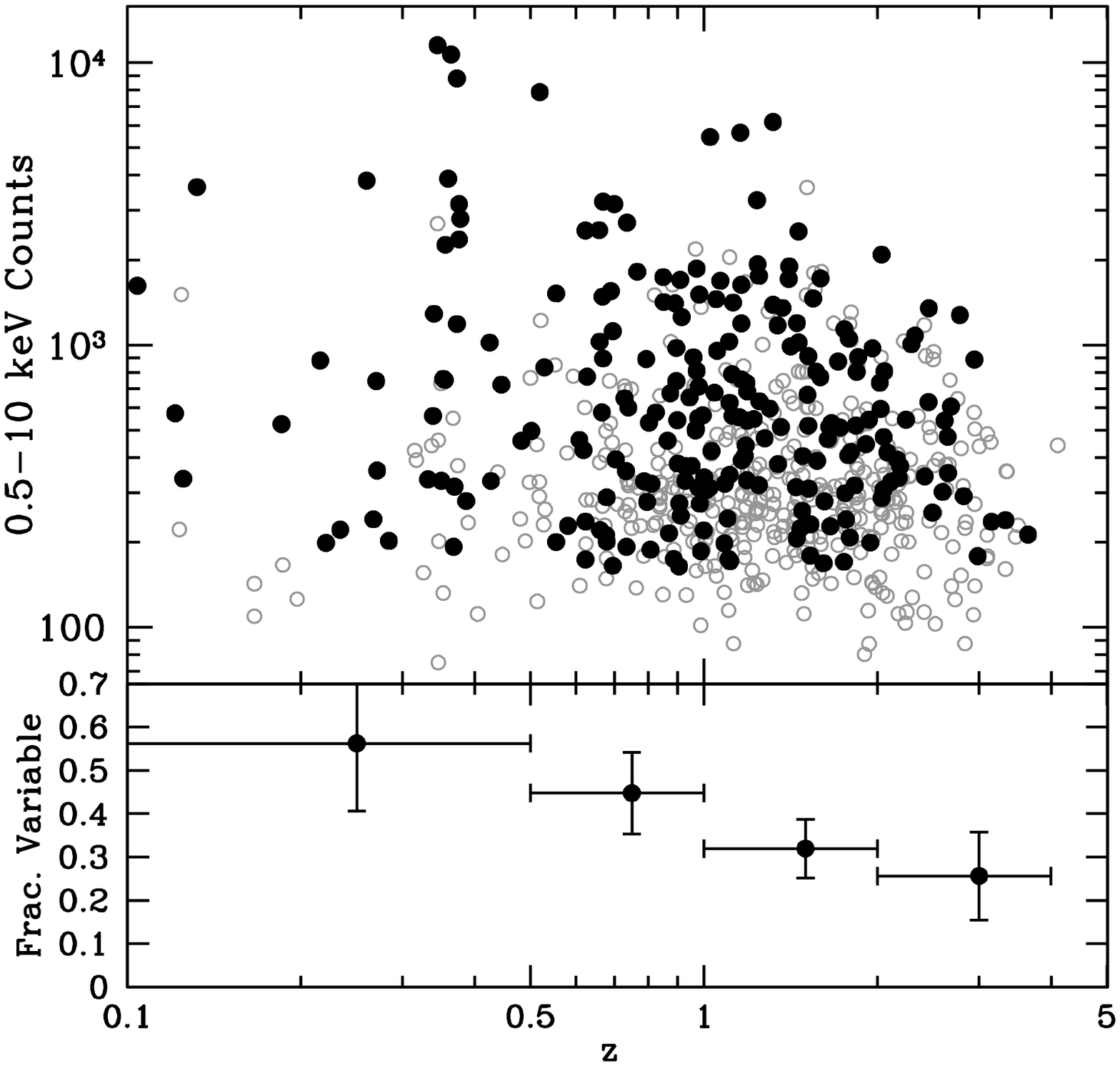}
\caption{{\it Left panel:} Distribution of 0.5-10 keV intrinsic luminosity as a function of 0.5-10 keV net counts (top).
Sources with $V > 1.3$ are labeled in black. The fraction of variable sources in bins of \lum\
is shown in the lower panel.
Vertical error-bars represent the Poissonian error on the measurement, horizontal error-bars show the bin size. 
{\it Right panel:} Same as left panel, but as a function of redshift.}
\end{center}
\label{lumcz}
\end{figure*}

\section{The V parameter}

To obtain an estimate of the probability of variability for each source, we adopted
the method described in McLaughlin et al. (1996) and used also
in Paolillo et al. (2004) and Young et al. (2012). 
To statistically quantify the variability of the sources, we computed
\begin{equation} \chi^2 = \sum_{i=1}^{N_{obs}} \frac{(x_i - \overline{x})^2}{\sigma_{err,i}^2}, \end{equation}
where N$_{obs}$ is the number of pointings in which the source was detected,
$x_i$ is the flux of the source in each pointing and $\sigma_{err,i}^2$ is the error, 
$\overline{x}$ is the average flux.
For intrinsically non-variable sources the value of the $\chi^2$ is
expected to be  $\sim N_{obs}-1$ (i.e. $\chi^2/d.o.f=1$). 
To determine whether the $\chi^2$ value is consistent with flux variability, we computed the
probability P($\chi^2$) that a $\chi^2$ lower than the observed could occur by chance,
for an intrinsically non-variable source. 
We then define the variability index V $= - log (1-P)$ to synthetically express the strength of evidence
of variability (i.e. V=1 correspond to 90\% confidence, V=1.3=95\%, V=2 to 99\% and so on).
A typical value for the critical probability P$_{crit}$, used to discriminate between variable and non-variable sources
in previous works with comparable sampling (but on smaller samples of AGN) is 95\% (Paolillo et al. 2004; Papadakis et al. 2008; Young et al. 2012).
Following this approach, we selected sources with $V>1.3$ (P$_{crit}=95\%$), i.e.  
sources that have only 5\% probability that the variability observed is due to Poisson noise alone and the source is intrinsically non-variable.
This translates into a sample of 232 variable sources, with an expected number of false positive of $\sim32$.
Adopting a more conservative approach, and selecting sources with $V>2$ (P$_{crit}=99\%$),
we would select a smaller sample of variable sources (164) with a smaller number of false positive ($\sim7$).
This would eliminate $\sim25$ false positive but it would also miss a similar amount of truly variable sources.
We stress, however, that we verified that using the sample selected with the more conservative approach ($V>2$), 
produces very similar results of the ones presented in the following, only in a smaller sample, and thus larger uncertainties.


\begin{figure*}
\begin{center}
\includegraphics[width=8cm,height=8cm]{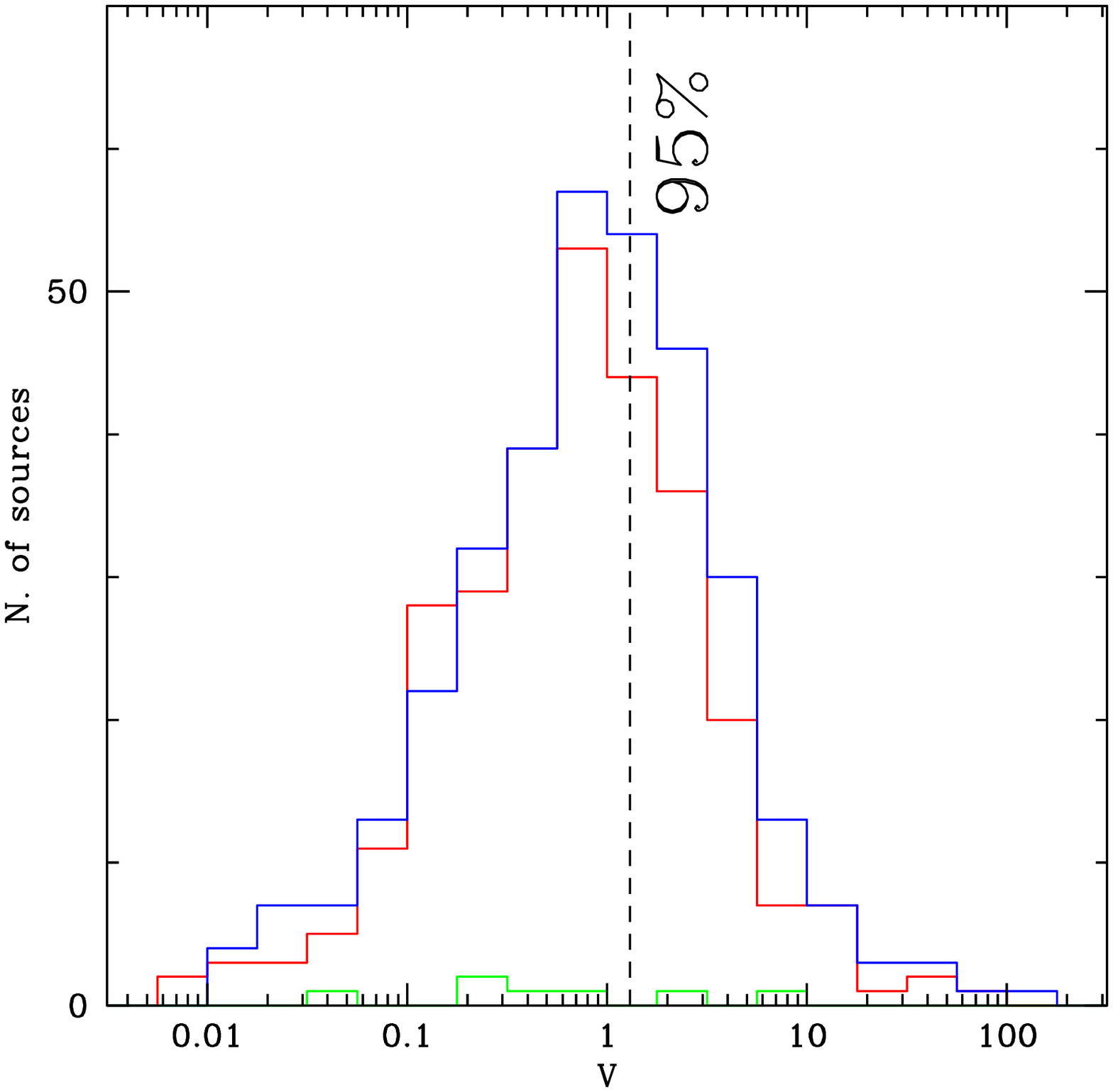}\hspace{0.5cm}
\includegraphics[width=8cm,height=8cm]{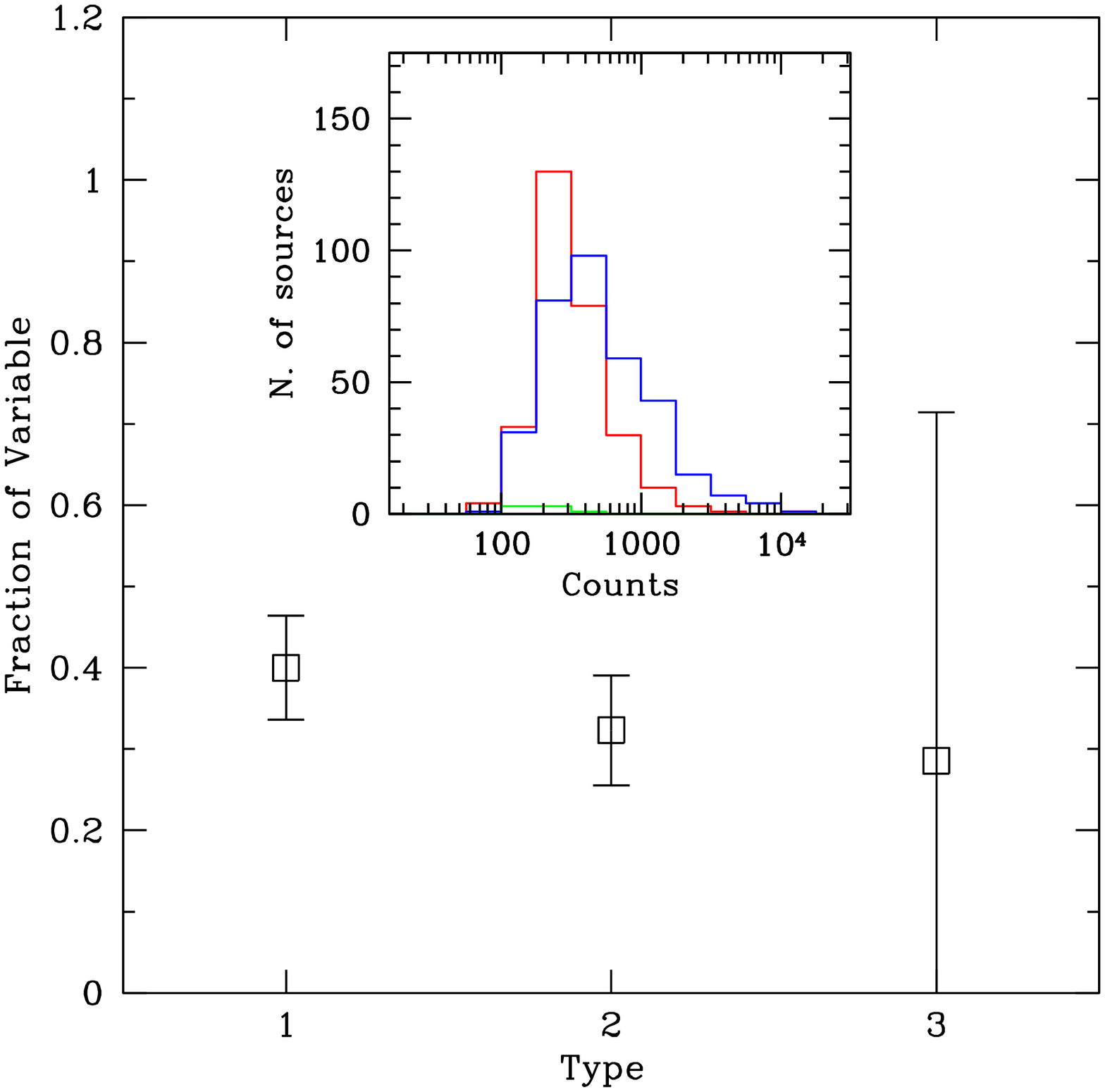}
\caption{{\it Left panel:} Distribution of the variability parameter V, for the 638 sources with available 
light curves in the 0.5-10 keV band,
divided by class. Type-1 AGN are shown in blue, type-2 AGN in red and Galaxies in green.
{\it Right panel:} Fraction of variable source (V$>1.3$) for the different classes: 1 for Type-1, 2 for Type-2 and 3 for galaxies.
The inset shows the distribution of counts per type of sources, colors as in the left panel.}
\end{center}
\label{istov}
\end{figure*}

\subsubsection{V vs. counts}

The V parameter is strongly dependent on the number of counts available, given that the ability to constrain 
the variability of one source depends on the error on each measurement. 
Fig. 2 (left panel) shows the distribution of the V parameter as a function of the total number
of counts for each source.
In Fig. 2 (right panel) we show the fraction of variable sources ($V>1.3$) in bins of 0.5-10 keV counts.
The fraction decrease from $\sim75\pm13\%$ above 1000 counts, to $\sim15\pm10\%$ in the last bin,
i.e. given enough statistic to detect it, variability is almost ubiquitous in X-ray selected AGN. 
However, even at very high number of counts, there is a small fraction of sources found to be non-variable with high reliability.

In Fit. 2 (left) 3 sources are highlighted, namely \xmm\ ID (XID) 3, 5323 and 2186. 
The X-ray and optical light curves for these sources
are shown in Fig. 3 upper and lower panel respectively (optical light curves from Salvato et al. 2013 in prep.)
Each one represents one extreme of the distribution:
source 3 has the highest number of X-ray counts (and the maximum number of detections: 9) 
and is highly variable, both in X-ray and optical;
source 5323 has again very good X-ray statistics, but is non-variable with very high confidence (only $5\%$ probability of being variable)
and is non-variable also in optical;
source 2186 has a more typical number of counts for our sample ($\sim200$ net counts in 0.5-10 keV) and is variable,
both in X-ray and optical.
 Source 5323 and 2186 have the same observed lightcurve duration, while source 3 was detected during the whole 3.5 years of observations.
However source 5323 is a luminous quasar at high z (log(L$_X$)=45.4 and z=1.509, with Log($M_{BH}$)=8.9), 
while 2186 is a low luminosity Seyfert at low redshift (log(L$_X$)=42.1 and z=0.235, with Log($M_{BH}$)=7.3). 
As we will show in Sec. 5, the different luminosity (and MBH) appears to have a  
major role in producing these different variability properties.

In total, 232 over 638 sources (36.4\% of the sample) exhibit significant flux variability on month-years rest-frame timescales.
We can compare this result with the ones of  Young et al. (2012) on CDF-S \chandra\ data.
They found 50\% of AGN to be variable.
This difference is probably due to the different background contribution for \chandra\ and \xmm: 
for a given total number of net counts, the S/N, and hence the error on the flux measurement is typically
lower (larger errors) for \xmm\ data.
The \chandra\ background accounts for only 5-10\% of source counts even for faint sources, 
while the background for faint sources in the \xmm-COSMOS survey can account to up to 50\% of source counts in the full band.

\subsubsection{V vs. \lum\ and z}

We are interested in the possible correlation between the X-ray variability and 
more intrinsic properties of the sources in our sample, such as \lum\ and redshift.
However, the strong dependency of V on number of counts could
produce biased correlations between these quantities.
We therefore explored the distribution of \lum\ and redshift as a function of the number of counts
(Fig. 4 left and right, top panels), for the whole sample (gray points) and for variable sources ($V > 1.3$,  black points).
The resulting fraction of variable sources in bins of \lum\ and z is show in the lower panels.

In Fig. 4 (left) there is a clear upper boundary in the distribution of the luminosities (\lum$\sim5\times10^{45}$ erg s$^{-1}$),
due to the angular size of the COSMOS survey and the low density of very luminous AGN, 
and a lower boundary in the number of counts ($\sim50$),
due to  the detection threshold of each observation, and the minimum number of detection of 3.
Furthermore, it can be seen that there is a general decreasing density of sources
going toward low luminosities, due to the smaller volume sampled at low z.
Finally there is a zone of missing sources, at low luminosities and high number of counts, 
due to the typical exposure time of the survey ($\sim40$ ks vignetting corrected). 
However, the fraction of variable sources (i.e. the relative distribution of variable and non-variable sources)
is constant, close to 40\%, in all the \lum\ range we considered ($10^{41}<$\lum$<5\times10^{45}$ erg s$^{-1}$).

The distribution of counts as a function of redshift is shown in Fig. 4 (right).
Here the zone of missing sources, at high redshift and high number of counts, 
is again due to the properties of the survey (exposure time and angular size):
there are no sources more luminous than $5\times10^{45}$ erg s$^{-1}$, and the almost constant
exposure time dictates a decreasing number of counts with z.
In this case however the final effect is a decrease in the fraction of variable sources
at high redshift, going from 55\% to 25\%  from low to high redshift. 
We stress that the distribution of the V parameter is an indicator of our ability 
to detect variability in our sample, and therefore Fig. 4 (right) gives an indication that, 
going at high z, we are less able to detect and measure variability
(see discussion in sec. 5).

\subsubsection{V vs. Optical type}

Fig. 5 (left panel) shows the distribution of the variability parameter V, for the 638 sources with available light-curve in the 0.5-10 keV band,
divided by class.
The dashed line represents the probability threshold we choose to define a source as variable.
Fig. 5 (right panel) shows the fraction of variable sources for the different classes.
 Type-1 AGN have the highest fraction of variable sources ($\sim40\%$),
type-2 have a lower fraction ($\sim32\%$), while galaxies have only a very large upper-limit
(only 1 variable source out of 7 in this class).
The inset of Fig. 5 (right) shows however that the higher fraction of variable type-1 (with respect to type-2 sources),
could be due to the typical higher number of counts available for these sources. 
Therefore, as observed also in the CDF-S (Young et al. 2012) variability is common in AGN samples, independently of the optical 
classification.

\section{The excess variance \sig}

To obtain a quantitative measure of the variability amplitude,
we calculated the normalized excess variance  
(\sig; Nandra et al. 1997; Turner et al. 1999) for all the 638 sources with more than 2 detections.
The normalized excess variance is defined as:

\begin{equation} \sigma_{rms}^2 = \frac{1}{(N_{obs}-1)\overline{x}^2} ~ \sum_{i=1}^{N_{obs}} (x_i - \overline{x})^2 - \frac{1}{N_{obs}\overline{x}^2} ~ \sum_{i=1}^{N_{obs}} \sigma_{err,i}^2, \end{equation}
where $N_{obs}$ is the number of observations, $\overline{x}$ is the average flux, 
$x_i$ is the single measurement with error  $\sigma_{err,i}$.
The $\sigma_{rms}^2$ measures how much of the total flux per observation is variable, after subtracting the statistical error.
Note that this form is usually simplified using the factor $1/N_{obs}$ in both terms, that
for a large number of observations gives consistent results.
In our case instead, given the typically small number of observations available, 
the difference between $N$ and $N-1$ is significant.

 The error on the excess variance due to Poisson noise is given by Vaughan et al. (2003; see also Ponti et al. 2004):
\begin{equation} 
err(\sigma^2_{rms})= \sqrt{ \left(\sqrt{\frac{2}{N_{obs}}} \cdot \frac{\overline{\sigma^2_{err}}}{\overline{x}^2}\right)^2 + \left( \sqrt{\frac{\overline{\sigma^2_{err}}}{N_{obs}}} \cdot \frac{2F_\mathrm{var}}{\overline{x}} \right)^2 }
\end{equation}
where $\overline{\sigma^2_{err}}$ is the mean square error: 
\begin{equation} 
\overline{\sigma^2_{err}}= \frac{1}{N_{obs}}\sum_{i=1}^{N}\sigma^2_{err,i}
\end{equation}
and $F_\mathrm{var}$ is the the fractional variability (Edelson, Krolik \& Pike 1990), $F_\mathrm{var}=\sqrt{\sigma_{rms}^2}$.

If the intrinsic excess variance is consistent with zero, due to statistical
fluctuations, the measured \sig\ can be negative. We consider it as
"detection" if \sig-err(\sig)$>0$, (there are 228 sources in our
sample with such "detections"). 
For the remaining 410 sources, we define an "upper limit" to the measured excess variance as: UL= \sig +err(\sig).
We underline that in the following we will use these UL for the fit with the survival analysis,
but we also tested that consistent results are obtained if $3\sigma$ UL are used instead.

It is well known that, apart from the Poisson noise, 
there are other sources of uncertainties in the determination of 
the excess variance, related to the intrinsic scatter of the variability,
and to the sampling of the light curves, which become even more important for sparsely sampled light curves.
Allevato et al. (2013) showed through Monte Carlo simulations that
the values of the excess variance, measured from either continuously or sparsely sampled
light curves, differ from the intrinsic, band normalized, excess variance.
In the case of  sparsely sampled light curves, the measured \sig\ is typically
an underestimate of the intrinsic \sig\ , because of the frequencies that are not well sampled.
The bias factor 
\footnote{Defined as the ratio between the intrinsic \sig\ and the \sig\ observed 
with a sparse sampling.} 
of a single measurement depends on the slope of the PSD, $\beta$, on the lightcurve sampling (continuous, uniform, sparse)
and on the signal to noise of each measurement.

In the case of the \xmm-COSMOS survey analyzed here, the lightcurve sampling is a worst case scenario 
as the fraction of time sampled is very small with respect to the total light curves length.
However, we are sampling the low frequency part of the PSD, where the PSD slope is supposed to be flat ($\beta\sim1$), and  
80\% of our sources have a total (0.5-10 keV) flux above $1\times10^{-14}$ erg cm$^{-2}$ s$^{-1}$.
Therefore the average bias factor is expected to be rather small ($\sim1.2$). 
Nonetheless, the standard deviation of the distribution of bias factors is large.
Therefore the bias for a specific measurement can be very far from 1 for a significant fraction of sources.
As a consequence, the \sig\ measured for a specific source should be taken cautiously. 
One way to overcome this problem is, however, to compute average \sig, for samples of sources with similar properties.
In this case, the scatter of the individual measurements around the average \sig\ 
can be used as the uncertainty on the average \sig.

\begin{figure}
\begin{center}
\includegraphics[width=8cm,height=8cm]{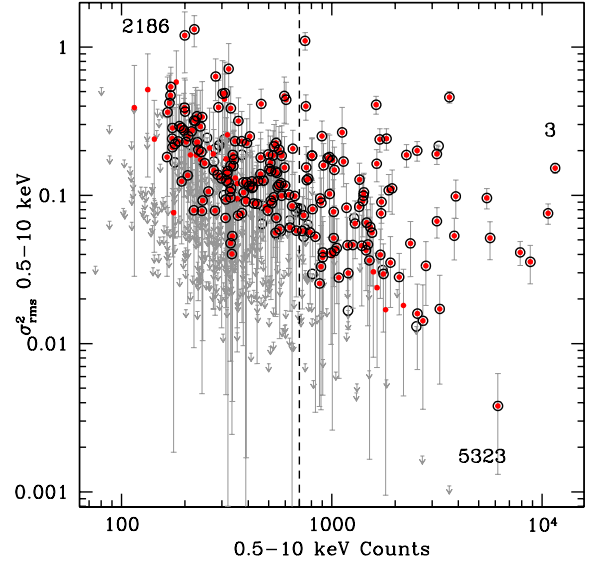}
\caption{Distribution of \sig\ as a function of total net counts for each source.
In red are labeled sources with detection on \sig. In gray, sources with an upper-limit.
Black circles mark sources considered variable on the basis of their V parameter.
The dashed line marks the 700 total net counts cut used in see Sec. 5.1.
The sources shown in Fig. 3 are labeled with their XID.}
\end{center}
\label{sigmac}
\end{figure}

\subsection{\sig\ vs. Counts}

Fig 6 shows the distribution of the \sig\ as a function of total net counts for each source.
In red are labeled sources with a detection on \sig. In gray, sources with an upper-limit.
Black circles mark sources considered variable on the basis of their V parameter ($V>1.3$).
The distribution of \sig\ is in Log scale, and upper-limits,  defined as \sig +\sig$_{Err}$, 
for all the sources with \sig-\sig$_{Err}<0$, are reported.
We note that all the upperlimits are positive (i.e. \sig +\sig$_{Err}>0$).

There is clearly an anti-correlation between \sig\ and  number of counts.
We underline that this anti-correlation
is produced by the increasing minimum upper-limit measurable for a given number of counts\footnote{
As for the variability index V, the ability to constrain the $\sigma_{rms}$, depends 
strongly on the signal to noise (and hence number of counts) of each source.
At $\simgt3000$ counts we are able to detect excess variance values as small as \sig=0.04, 
while, at lower counts, the value of \sig\ must be much higher to be detected (e.g. $>0.3$ at $\sim50$ counts).
The same is true for the minimum upper-limit value, that goes from $\sim0.001$ above 3000 counts, to $\sim0.1$ below 100 counts.},
and hence the increasing minimum level at which we can have a detection,
and by the fact that we have very few sources with high number of counts,
and hence the region with high \sig\ and high number of counts is less populated.
The dashed line corresponds to the cut of 700 total net counts, that we will use in Sec. 5.1 to evaluate the effect of the selection in V.
We underline that above this line there is instead no correlation between \sig\ and number of counts,
unlike what found for the total sample.

\begin{figure*}[t]
\begin{center}
\includegraphics[width=8cm,height=8cm]{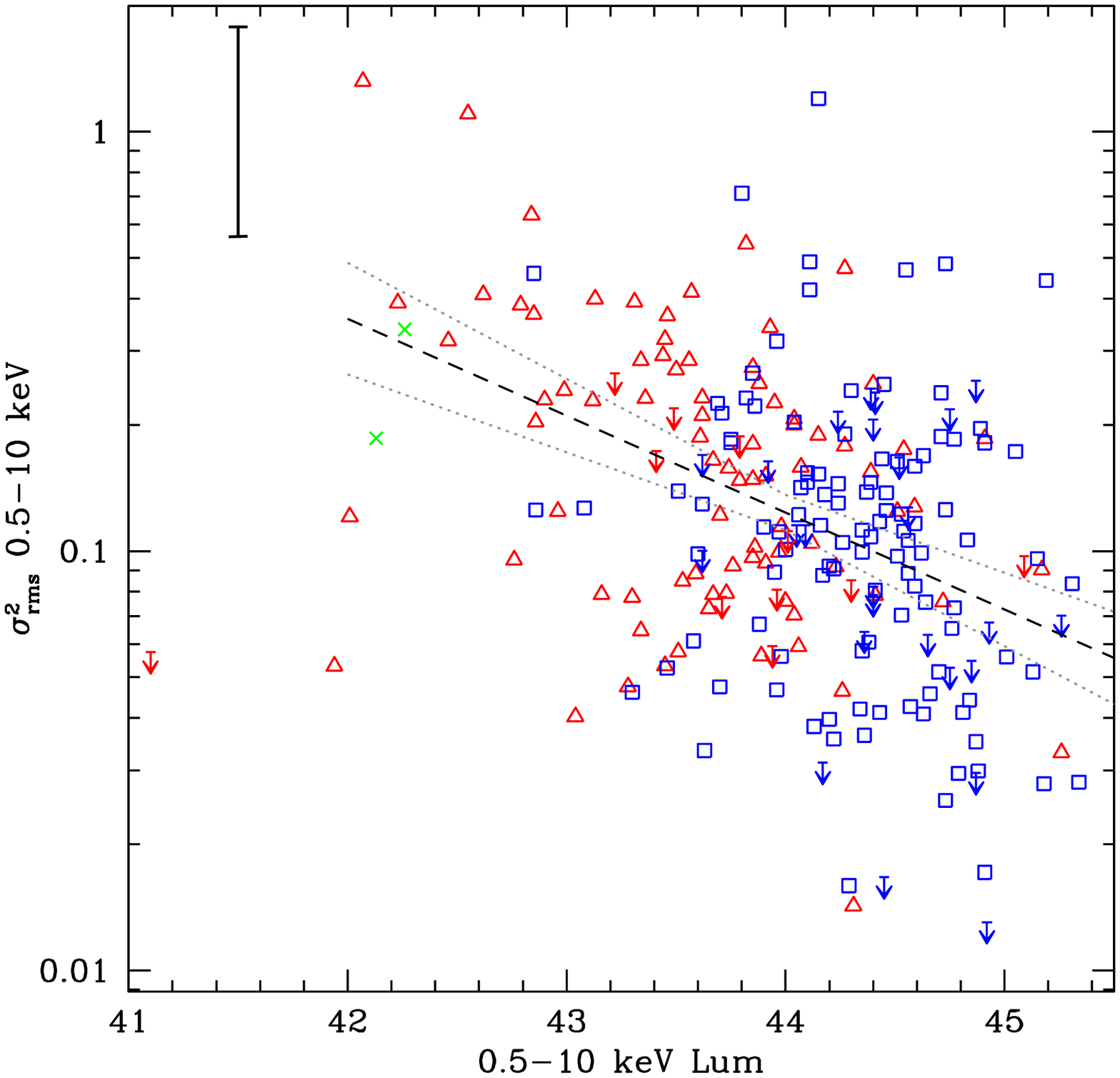}\hspace{1.cm}
\includegraphics[width=8cm,height=8cm]{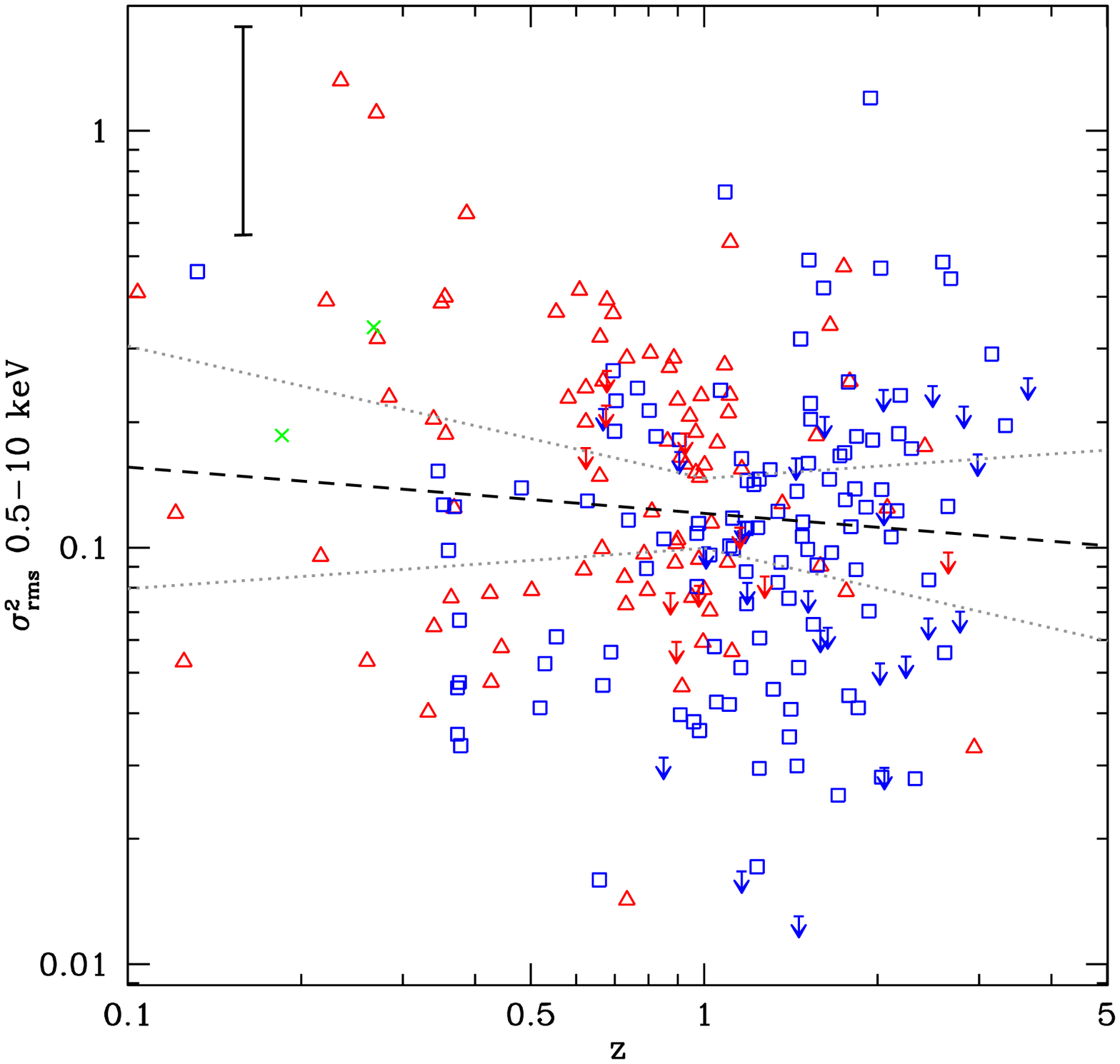}
\caption{{\it Left panel:} Distribution of $\sigma_{rms}$ as a function of the 0.5-10 keV band intrinsic luminosity 
for the sample of sources with $V>1.3$. 
{\it Right panel:} Distribution of $\sigma_{rms}$ as a function of redshift, for sources with $V>1.3$.
In both panels the dashed line shows the linear regression, including upper-limits, for the total sample. 
The dotted lines show the errors in the coefficients. The black solid error-bar shows the average error on \sig.
Blue squares are type-1 sources, red triangles are type-2, green crosses are galaxies.
}
\end{center}
\label{ltstz}
\end{figure*}

If we assume that a detection in \sig\ can be considered as indication of intrinsic variability,
it is interesting to compare the distribution of \sig\ detections with that of sources sources showing $V>1.3$.
We recall however that the two methods express different concepts:
that the V parameter indicates at which confidence a source can be considered variable,
while the excess variance measures the amplitude of the variability.
Nonetheless the two methods give very similar results:
There are only 21 sources detected in \sig\ but non variable according to the V parameter,
and 36 sources variable in V but not detected in \sig.
The agreement is also visually represented in Fig. 6, where there is a high correspondence of 
red dots (detections in \sig) and black circles ($V>1.3$ sources).

\section{\sig\ vs. AGN physical properties}

Due to the generally limited photon statistics of XMM-COSMOS,
the majority of the sources in our sample have an upper-limit on \sig\ because
of the large errors associated with the small number of counts.
We want to study how variability (namely the \sig) correlates with 
other quantities, such as \lum, and z, BH mass, Eddington ratio and optical properties,
for sources for which we are able to constrain this quantity.
Therefore in the following, we will first discuss such 
correlations for the subsample of 232 sources found to be significantly variable ($V>1.3$), 
such as done in previous works (Paolillo et al. 2004; Papadakis et al. 2008, Young et al. 2012).
However, we will also show that this selection criterion produces a strong bias in the distribution of \sig,
while instead selecting sources on the basis of the available number of counts produces more reliable results.
We also note that the distributions of \sig\ for type-1 and for type-2 sources are completely consistent. 
Therefore we will study the distribution of \sig\ as a function of other physical parameters for 
type-1 and type-2 together.

\subsection{\sig\ vs. \lum\ and z}

\begin{figure*}
\begin{center}
\includegraphics[width=8cm,height=8cm]{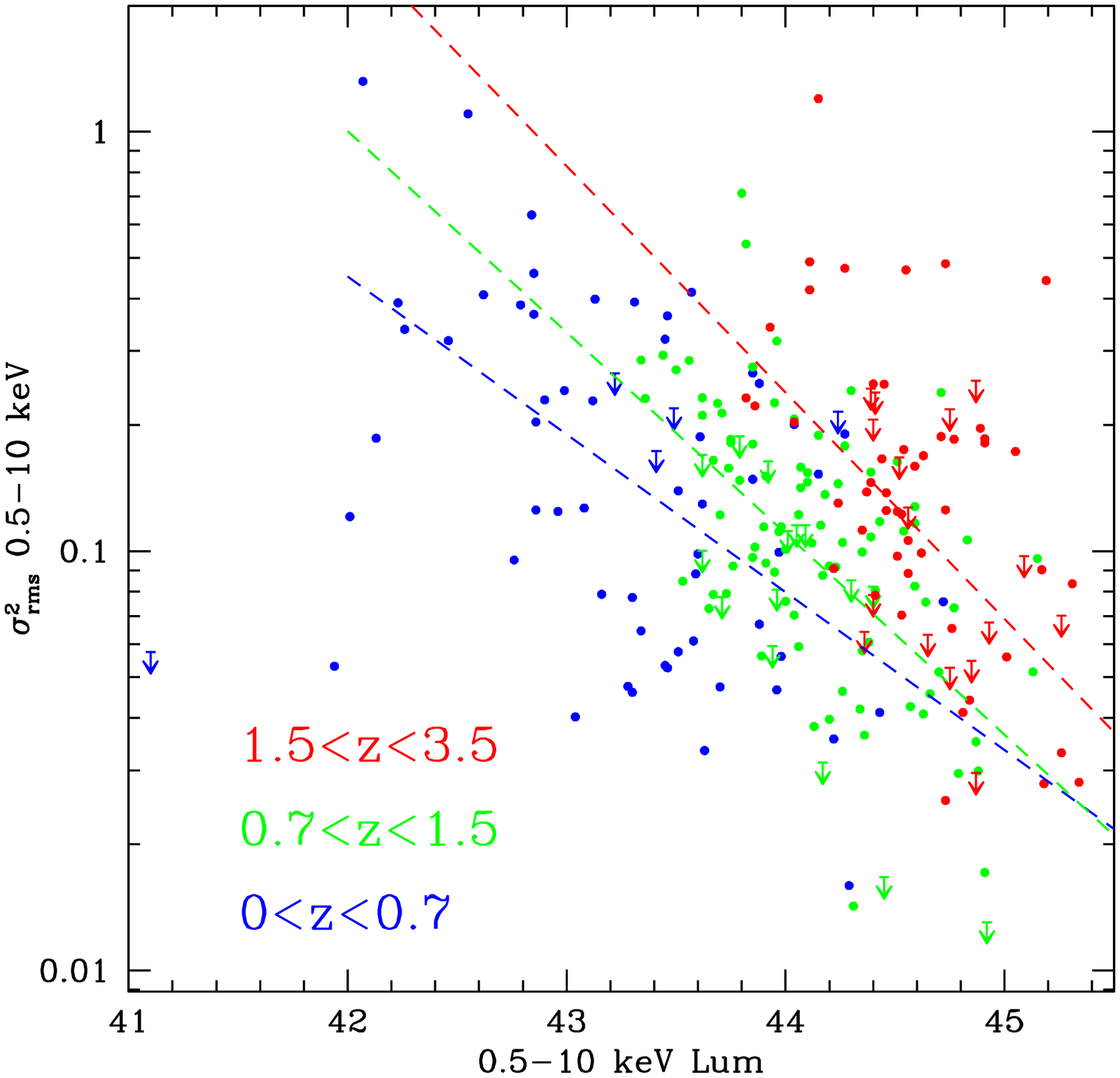}\hspace{1.cm}\includegraphics[width=8cm,height=8cm]{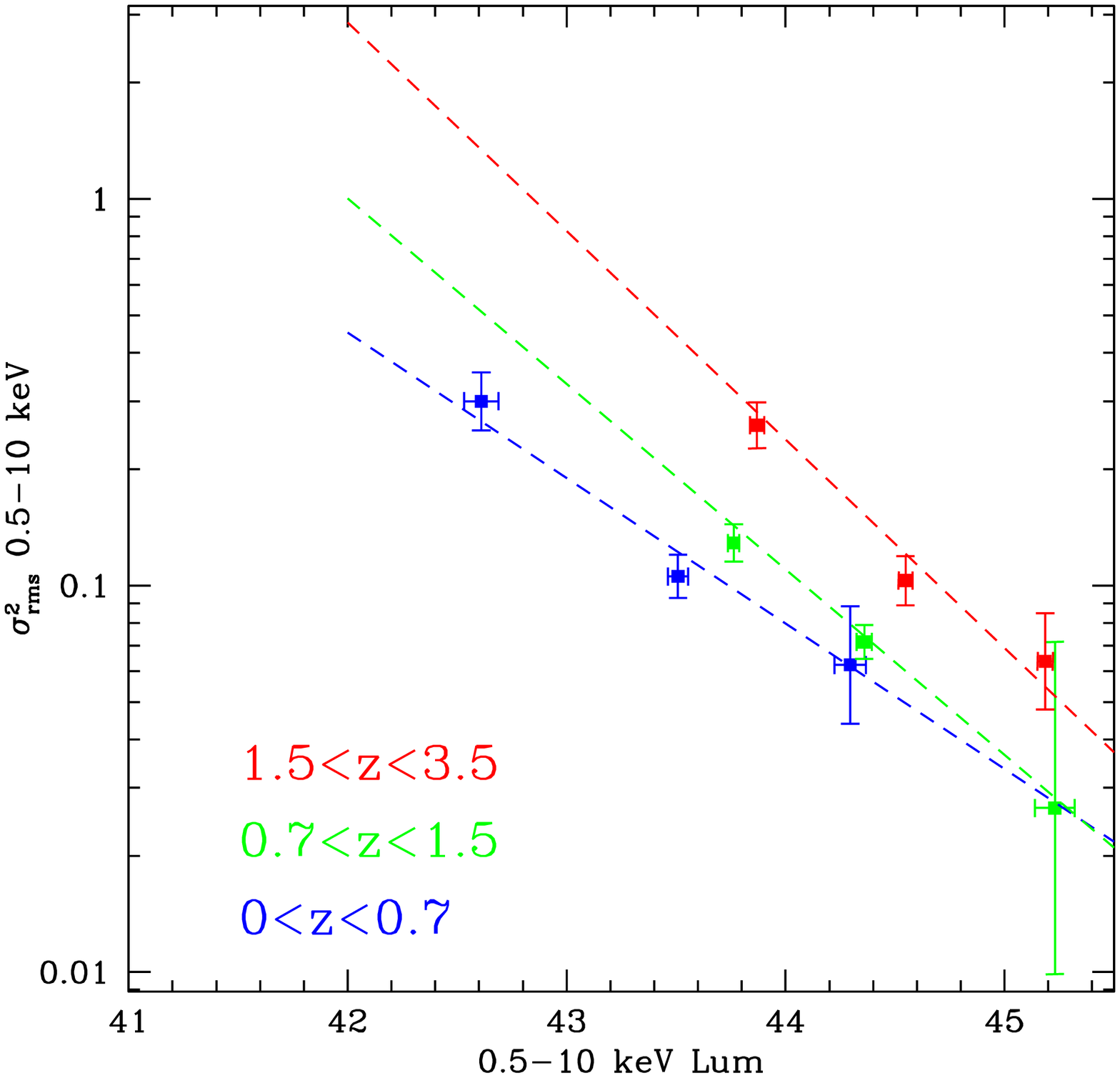}
\caption{{\it Left panel:} Same as Fig. 7 (left), but here the sample has been divided in 3 redshift bins ($z<0.7$, $0.7<z<1.5$, $1.5<z<4.5$).
Sources and linear regressions are color-coded so that blue represent the lower redshift sample, 
green the intermediate, and red the high redshift one.
{\it Right panel:} For each z subsample, the average \sig\ and \lum\ are shown in bins of luminosities.
The squares mark the average of \sig\ and \lum\ for the bin, while the error-bars represent the standard error on the average.
The same linear regressions obtained for left panel is shown for reference.
}
\end{center}
\end{figure*}

Fig. 7 (left) shows the distribution of $\sigma_{rms}$ as a function of the 0.5-10 keV band
luminosity for the sample of 232 sources with $V>1.3$. 
The X-ray luminosities are corrected for absorption, either from spectral analysis (Mainieri et al. 2007)
or hardness ratio (Brusa et al. 2010), and computed in the 0.5-10 keV rest frame band.
Blue squares are type-1 sources, red triangles are type-2, green crosses are galaxies.
The dashed line represents the linear regression performed with the ASURV software  
Rev 1.2 (Isobe \& Feigelson 1992), which implements the
methods presented in Isobe, Feigelson and Nelson (1986).
 We adopted the results from the Buckley-James method, given that in all cases 
the censorship is present in only one variable. The results from the EM algorithm method are always consistent with 
the former ones, while results from the Schmitt's binned linear regression is not suitable, due to the limited number of data points.
We excluded from the fit the 2 sources at \lum$<10^{42}$ erg s$^{-1}$, that can have significant contamination 
by the host galaxy.
An anti-correlation between excess variance and X-ray luminosity
is observed with high significance: the Spearman's rank correlation coefficient ($\rho_S$=-0.375)
gives a probability of $P_S<10^{-4}$.
The resulting correlation is:

\begin{equation} Log(\sigma^2_{rms})=(-0.91\pm0.02)+(-0.23\pm0.03)Log(L_{2-10 keV,44}) \end{equation}

with \lum\ in units of $10^{44}$ erg s$^{-1}$.

A steeper anti-correlation for AGNs has been observed in previous work
both for high frequencies
(e.g., Barr \& Mushotzky 1986; Lawrence \& Papadakis 1993; Nandra et al. 1997; Hawkins 2000)
and on longer time scales (Paolillo et al. 2004;
Papadakis et al. 2008; Young et al. 2012).
In particular, Paolillo et al. (2004) found a slope of $-1.31\pm0.23$ in the CDF-S,
limiting the analysis to sources in the redshift range $0.5<z<1.5$.
Papadakis et al. (2008) found a flatter slope of $-0.66\pm0.12$ in the Lockman Hole
while Young et al. (2012) found a slope of $-0.76\pm0.06$ in the 4 Msec CDF-S data set
(for sources with \lum$>10^{42}$ erg s$^{-1}$), but not including sources with upper-limits.
Again the latter sample is limited to $z\leq1$.

Fig. 7 (right) shows the distribution of \sig\ as a function of z for the sample.
No significant correlation is observed ($\rho_S$=-0.031 and $P_S=0.6798$).
This is somewhat unexpected, because at high redshift we are sampling the high luminosity
part of the population, and therefore we would expect to observe a decrease in the typical \sig\ with z.
This is probably balanced by the fact that at high z we are selecting a smaller fraction of variable sources
(see Fig. 4, right), thus selecting only the most variable ones.

The large redshift and \lum\ range encompassed by the XMM-COSMOS survey, allows us to study more in detail this issue:
Fig. 8 (left) shows the distribution of \sig\ vs. \lum\ for sources divided in 3 redshift bins ($z<0.7$, $0.7<z<1.5$ and $1.5<z<3.5$). 
The dashed lines are linear regressions computed with ASURV for each subsample.
Sources and lines are color coded so that blue represent the lower redshift sample, 
green the intermediate, and red the high redshift one.
Dividing the population in this way, the anti-correlation between \sig\ and \lum\ is stronger:
each correlation has  $P_S<10^{-4}$ and the slopes go
from  $-0.40\pm0.07$ to $-0.65\pm0.10$ from low to high redshift.
Furthermore, it appears that, for a given luminosity range, and at all luminosities, 
the typical excess variance gets higher at higher redshifts.
The intercept coefficients are significantly different,
going from  $-1.21\pm0.07$ to $-0.48\pm0.07$ from low to high redshift.
An hint of a similar effect was observed in Paolillo et al. (2004),
but the smaller sample size and \lum-z coverage did not allow a robust investigation in that work.

\begin{figure*}[t]
\begin{center}
\includegraphics[width=8cm,height=8cm]{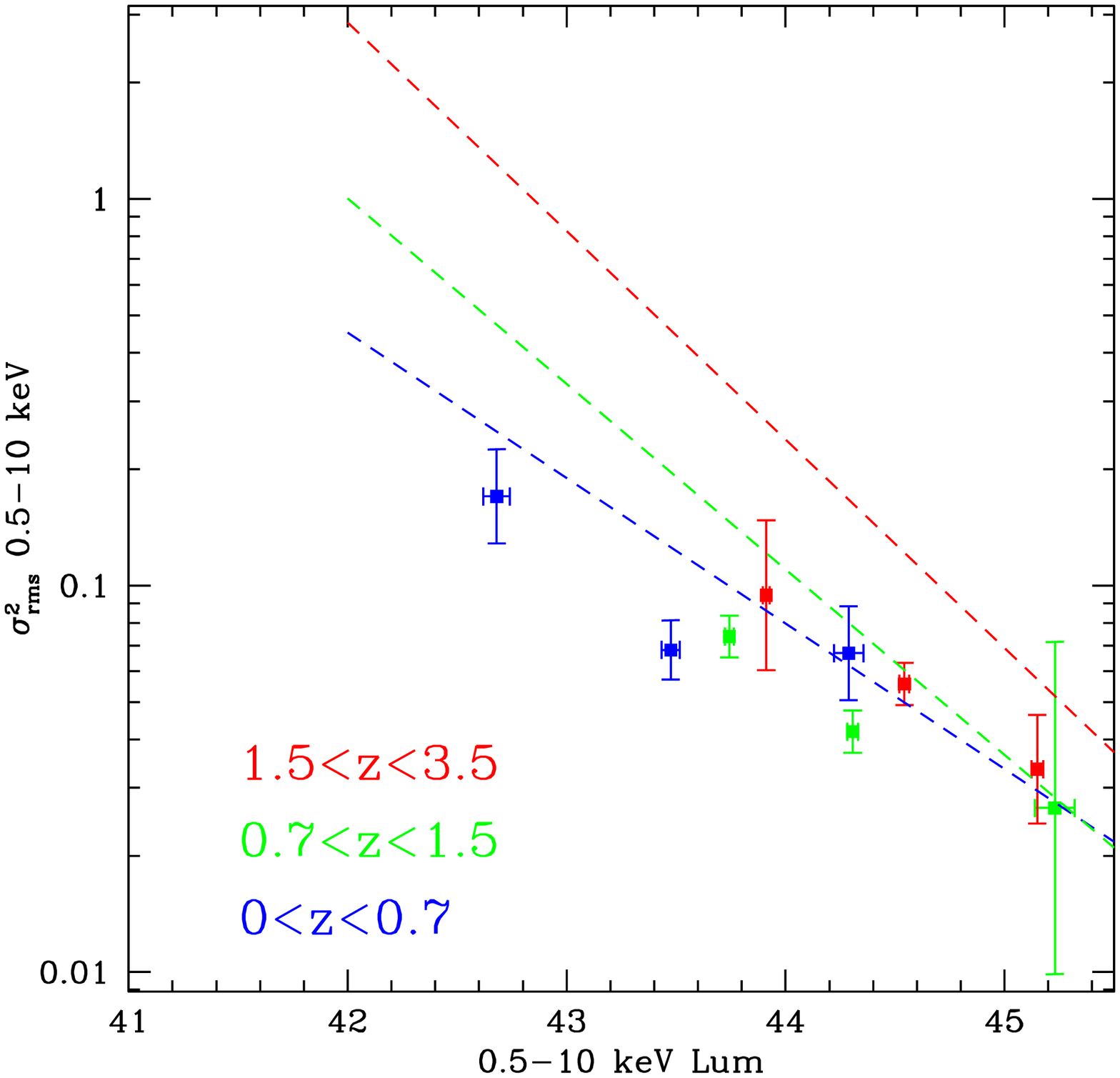}\hspace{1cm}
\includegraphics[width=8cm,height=8cm]{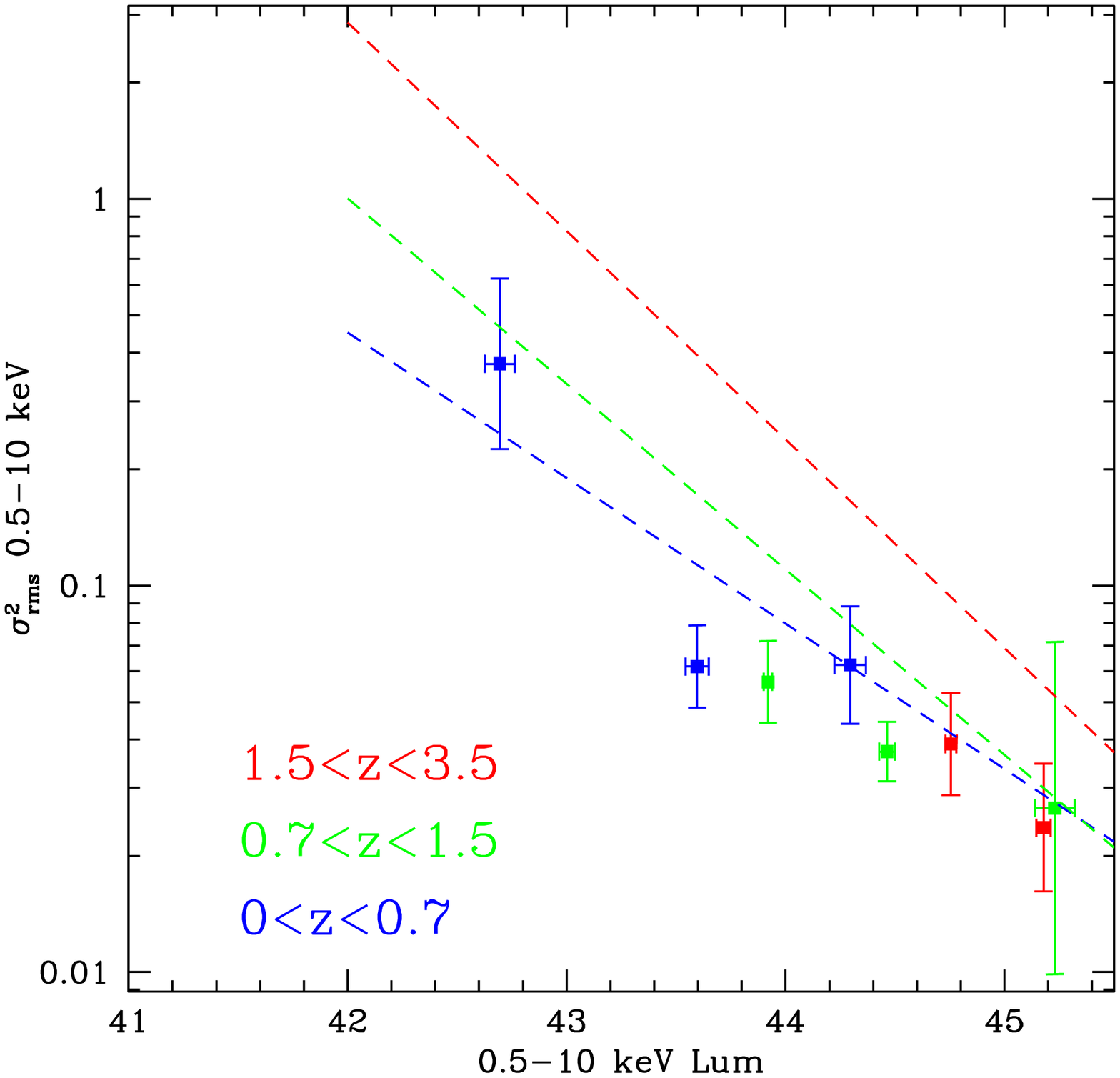}
\caption{{\it Left panel:} \sig\ vs. \lum\ average distribution, in redshift and \lum\ bins,
for all the 638 sources in the sample.
The squares mark the average of \sig\ and \lum\ for the bin, while the error-bars represent the standard error on the average.
{\it Right panel:} Same as top panel, but for sources selected 
on the basis of the number of counts available ($>700$ total net counts).
In both panels the same linear regressions obtained for Fig. 8 is shown, for reference.
}
\end{center}
\label{selection}
\end{figure*}

As explained in sec. 4, due to the broad distribution of the bias introduced by measuring 
the excess variance from sparsely sampled light curves, 
the \sig\ measured for a specific source can be far from the intrinsic one.
To overcome this problem, we computed the average excess variance for 
subsamples of sources showing similar physical properties (such as \lum, z, $M_{\rm BH}$ and Eddington ratio).
Fig. 8 (right) shows the distribution of average \sig vs \lum, for sources in the same redshift bins
of left panel, and further binned in luminosity bins (one bin per decade).
The position of the squares mark the average of \sig\ and \lum\ for the bin.
The average is computed taking into account all the values of \sig\ at face value
for both detections and upper-limits (e.g. the computed value regardless of the errors, that for upper-limits can be negative).
Error-bars represent the standard error on the mean for both quantities, in each subsample.
The linear regressions of left panel are shown for reference. 
We stress that the the linear regressions computed with ASURV for the distribution 
of points in Fig. 8 (left), are in perfect agreement with the distribution of the binned points.

We note that our decision to consider only the subsample of significantly
variable sources ($V>1.3$) avoids the difficulty of dealing with sparse 
light curves of faint  sources for which the Poissonian noise is far greater 
than the intrinsic variability. 
However, we note that the $\langle$ \sigv $\rangle$ averaged over the subset 
of the sources with $V>1.3$ is not the same as the $\langle$ \sig $\rangle$ averaged 
over the whole set of sources, and $\langle$ \sigv $\rangle$ will always 
be $\geq\langle$ \sig $\rangle$.
Therefore the selection $V>1.3$ introduces a further bias in the averaged excess 
variances. The bias (in the estimation of the mean excess variance) is stronger the larger
the number of faint sources in each luminosity and z bin that are left out.

From the light curves, in fact, we measure the variance ($S^2$) that 
has a contribution due to the intrinsic variability (\sig) and the 
Poissonian noise ($\sigma^2_{Poiss}$), \sig $= S^2 + \sigma^2_{Poiss}$ (Edelson et al 2002). 
If, as in the majority of faint AGN in XMM-COSMOS, 
$S^2 \sim \sigma^2_{Poiss} \gg $ \sig,
the uncertainty in the determination of the true value of $\sigma^2_{Poiss}$ will lead to either an over- or under-estimation 
of the excess variance, thus sometimes producing negative excess variances. 
The average of the excess variance estimates (considering both positive and negative values) is unbiased 
and it does tend to the real excess variance. On the other hand if, such as done here, we average only the variable 
sources ($V > 1.3$), thus preferentially discarding the negative excess variance values,
the obtained average excess variance will be overestimated, especially for those bins where  
a large fraction of faint sources are excluded.
This is why we attribute the increasing observed AGN variability at high redshift (see Fig. 8)
to this effect.
In fact, as shown in sec. 3.1.3, the fraction of variable source, selected with a fixed values of V, decreases with redshift. 
This means that, going at high redshift, there is an increasingly large number of sources that are left out 
from the bin.

To test this, and check how much of the observed effect is due to selection, 
we computed the same distribution of Fig. 8 (right), in two cases.
First (Fig. 9, left), we included in the average distribution, all the 638 sources with a lightcurve,
without any pre-selection. The linear regressions obtained in different redshift bins from Fig. 8,
are shown for reference.
The average \sig\ for each bin is, as expected, lower than before, because we are including in the average a large number
of sources with \sig\ close to 0 (positive or negative), and, as expected, the difference increases going at high redshift,
because more sources were excluded at high redshift by the $V>1.3$ selection (see Fig. 4 right).
Using the whole sample, the evolution in redshift is not significant any more.

\begin{figure*}[t!]
\begin{center}
\includegraphics[width=8cm,height=8cm]{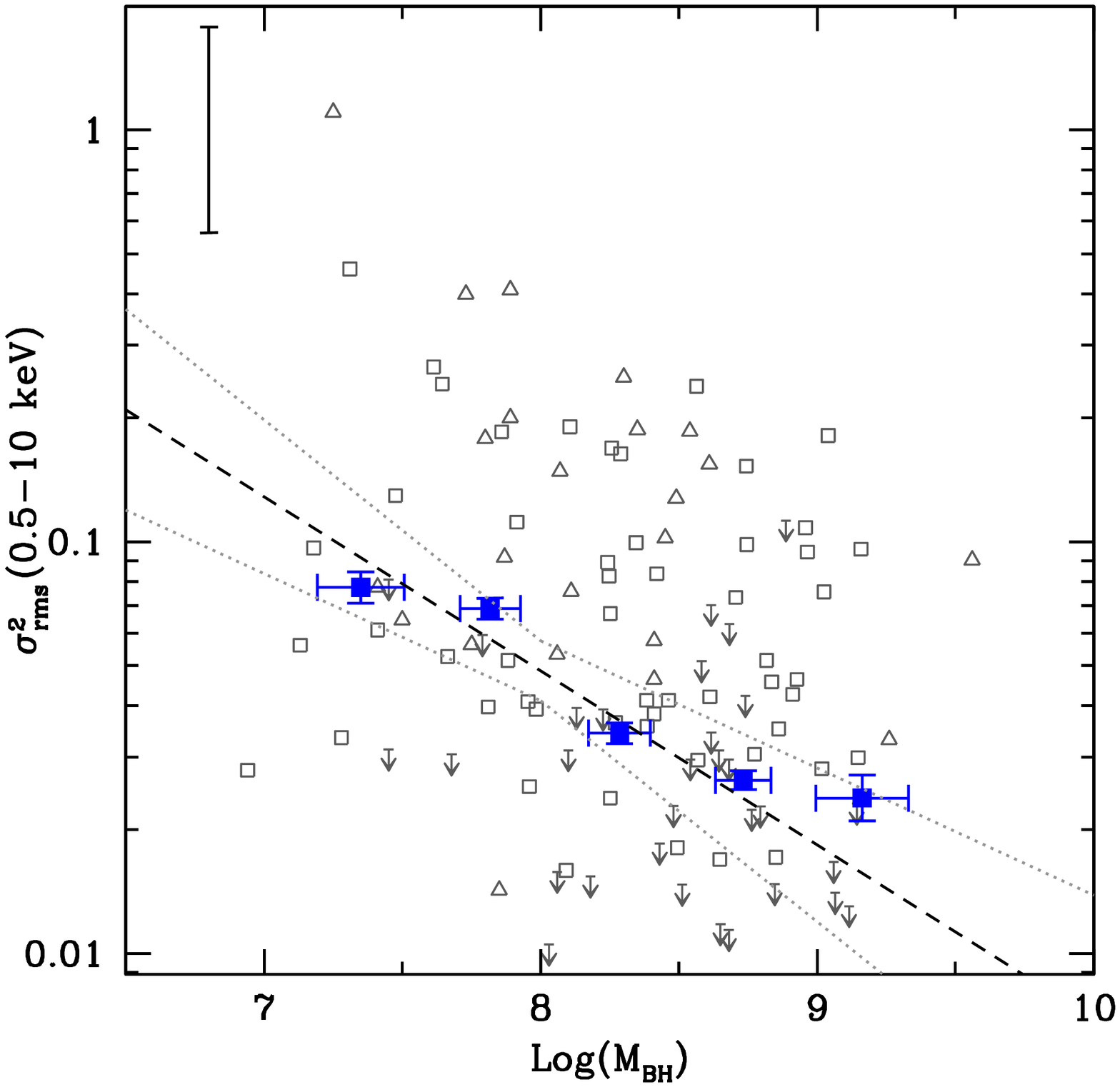}\hspace{0.5cm}
\includegraphics[width=8cm,height=8cm]{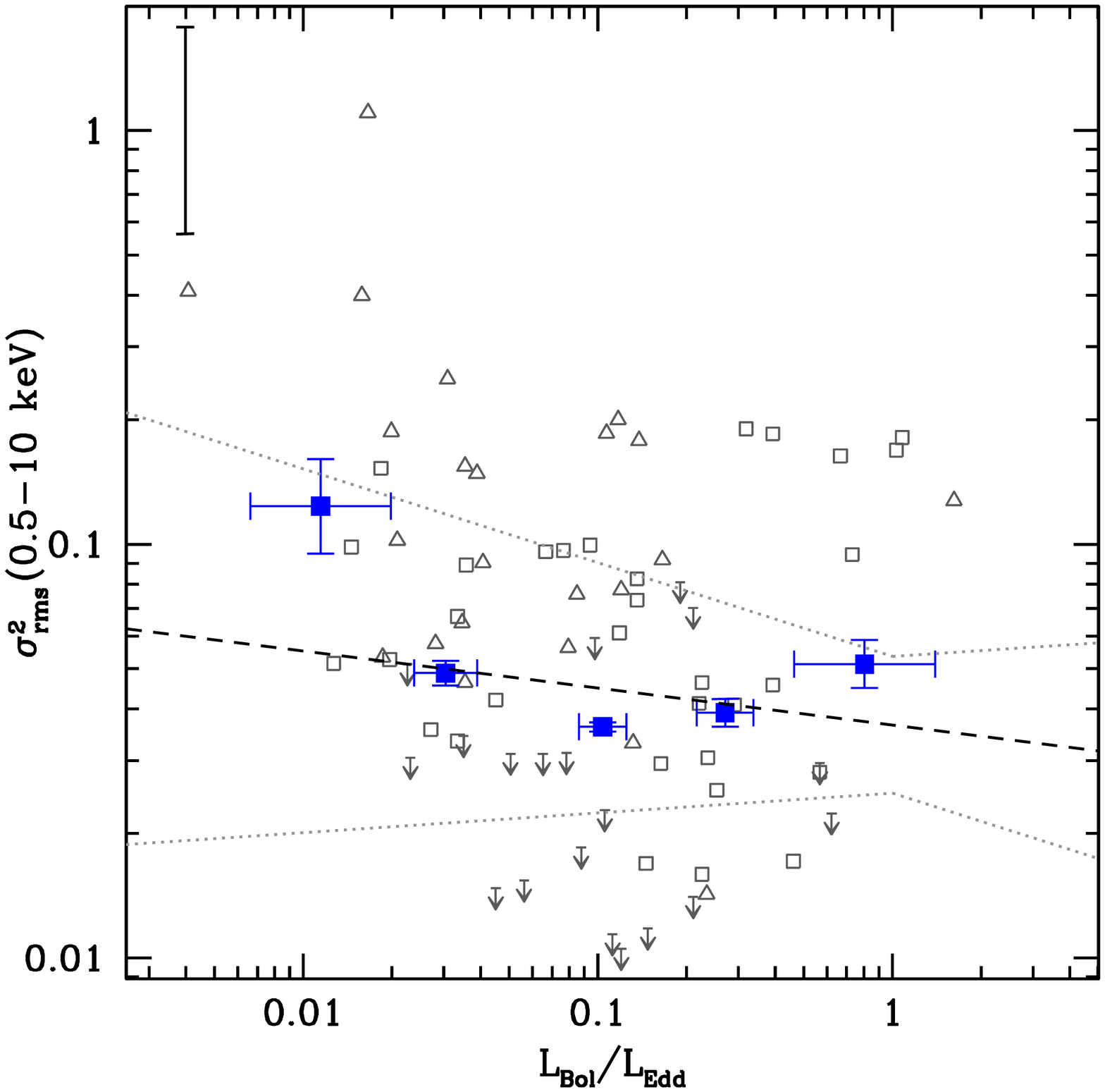}
\caption{{\it Left panel:} Distribution of \sig\ as a function of the BH masses, 
for sources with more than 700 total net counts, and no selection in the V parameter.
{\it Right panel:} Distribution of \sig\ as a function of Eddington ratio, expressed as L$_{Bol}$/L$_{Edd}$, for the same sample. 
In both panels the dashed line shows the linear regression, including upper-limits. 
The dotted lines show the errors in the coefficients. The black solid error-bar shows the average error on \sig.
Blue squares are type-1 sources, red triangles are type-2, green crosses are galaxies.
}
\end{center}
\label{sigmambh}
\end{figure*}

Finally, we showed in Sec. 4 (and Fig. 6) that the scatter in \sig\ increases going toward low counts, due to
poor statistics lightcurves. We want to test the effect of this, in the distribution of \sig\ and luminosity.
Therefore, in Fig. 9 (right), we selected sources on the basis of the number of counts available, i.e. sources with good statistic
($>700$ total net counts,  144 sources), for which we consider the estimation of variability more reliable.
No selection in V is used in this plot.
Also in this case all points are consistent with the local relation, i.e. the one measured for $z<0.7$:
\begin{equation} Log(\sigma^2_{rms})=(-1.21\pm0.06)+(-0.40\pm0.07)Log(L_{2-10 keV,44}) \end{equation}
even if the number of sources available for bin is small, and we are missing the low luminosity-high redshift bins.
{We underline that, even if the subsample used in Fig. 9 (right) is much smaller than the one in Fig. 9 (left),
(144 sources instead of 638), the smaller scatter produce similar error bars.}

These tests indicate that the selection based on the V parameter introduces a bias, at high redshift,
on the average excess variance.
Once we average all the excess variance values (with no selection in V),
the evidence for higher AGN variability at high redshift is not significant anymore for the XMM-COSMOS AGN.



\subsection{$\sigma_{rms}$ vs. $M_{\rm BH}$ and L$_{Edd}$}

We collected all the BH mass information available to date in the XMM-COSMOS survey.
For type-1s we have 89 virial masses available from MgII from Merloni et al. (2010);
128 from MgII, 31 from H$\beta$ and 37 from CIV from Trump et al. (2009);
69 from H$\alpha$ and 183 from MgII from Matsuoka et al. (2013).
We also have a recompilation of these type-1 BH masses, using the same spectra, 
but with a self consistent re-analysis, following Trakhtenbrot \& Netzer (2012). 
The different BH masses values, for sources with multiple measurements, are in general 
in good agreement, within 0.3 dex.
For type-2 we have 481 mass estimates through scaling relation and SED fitting from Lusso et al. (2011).
In total we have at least 1 mass estimate for 814 sources.

As we already showed in Sec. 5, the less biased selection to study the dependency of \sig\ from other 
physical properties in our sample, requires to select all sources (with no pre-selection in V).
 Furthermore, we decide to rely on the subsample of sources with  good enough statistic ($>700$ counts),
for which the estimation of variability is more reliable.
Using the same selection, we built a sample of 111 sources with $M_{\rm BH}$ estimates.
We stress that this is the first time in which the correlation between X-ray variability on long time scales
and BH masses can be performed, in such large sample of AGN, spanning a wide range of redshift.

The distribution of excess variance as a function of the BH mass is shown in Fig. 10 (left).
Single sources are shown in gray (squares for type-1 and triangles for type-2). 
The black error-bar shows the average (representative of a single measurement) error on \sig.
The position of the filled blue squares mark the average of \sig\ and $M_{\rm BH}$ in 5 bins of BH mass, 
where also sources with upper-limits are taken into account with their nominal value.
Error-bars represent the standard error on the mean for both quantities, in each subsample.

When only sources with good statistics are considered, 
the correlation between \sig\ and $M_{\rm BH}$ is strong ($\rho_S$= -0.315 and $P_S= 0.0007$),
with a slope of $-0.42\pm0.11$.
We note that the correlation between \sig\ and the $M_{\rm BH}$ found in previous work sampling higher frequencies (minutes-hours)
is of the order of -1 (e.g. P12). This is generally interpreted with the fact that
the \sig, measuring the integral of the PSD in that specific frequency range 
is affected by the position of $\nu_b$, that scale linearly with the BH mass.

 The lower frequencies sampled in the XMM-COSMOS survey, are in the months-years regime.
On the other hand, the highest frequency sampled by a sparsely sampled lightcurve is not well defined. 
However, in most lightcurves the minimum distance between two observations is of the order of hours-days.
Therefore we are integrating the PSD both above and below $\nu_b$. 
The part of the PSD integral above $\nu_b$ would introduce a linear correlation of 
the excess variance with $M_{\rm BH}$, but this is weakened by the part of the integral below $\nu_b$
that would induce no correlation (see for example Soldi et al. 2013). 

We also collected the AGN Bolometric luminosities from Lusso et al. (2010) for type-1 
(estimated from direct integration of the rest-frame SED), and Lusso et al. (2011) 
for type-2 (from SED-fitting by assuming a fixed covering factor of 0.67).
Thanks to this, we were able to compute Eddington ratios for 74 sources with $>700$ counts.
Fig. 10 (right) shows the distribution of \sig\ as a function of Eddington ratio
expressed as L$_{Bol}$/L$_{Edd}$.
As in Fig. 10 (left), gray points show single sources, blue filled squares 
represent the average of \sig\ and L$_{Bol}$/L$_{Edd}$ in 5 bins of Eddington ratio.
There is no e correlation between \sig\ and the Eddington ratio
($\rho_S$= -0.122 and $P_S=  0.2784$, with a slope fully consistent with 0, $-0.09\pm0.14$).

\begin{figure}
\begin{center}
\includegraphics[width=8cm,height=8cm]{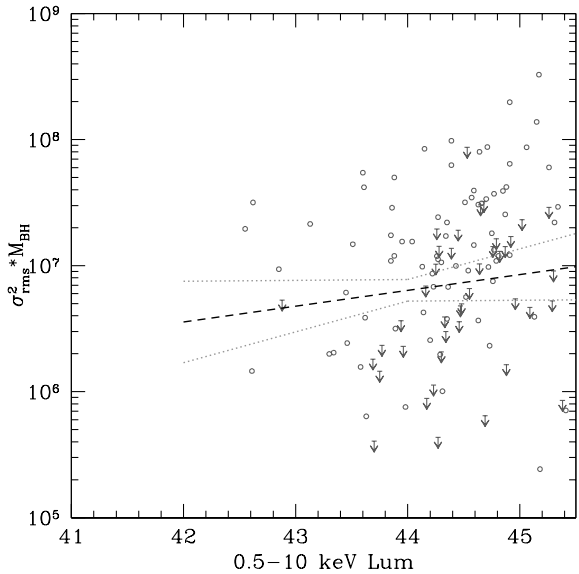}
\caption{Distribution of \sig * $M_{\rm BH}$ for sources  with more than 700 total net counts.
The linear regression computed for \sig\ normalized for the BH mass is consistent with 0.
}
\end{center}
\label{ltstmbh}
\end{figure}

\begin{figure*}[t]
\begin{center}
\includegraphics[width=8cm,height=8cm]{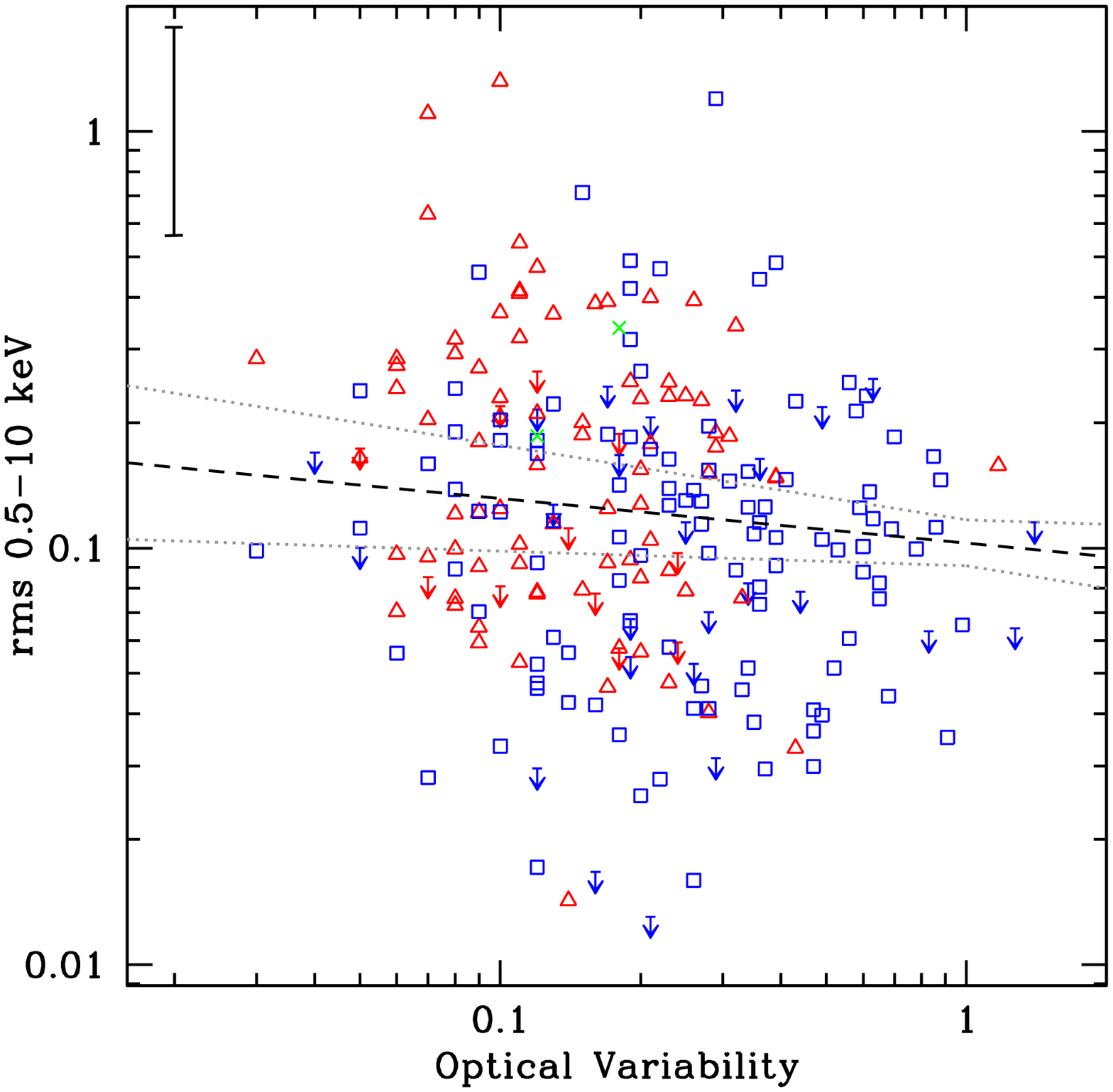}\hspace{0.5cm}
\includegraphics[width=8cm,height=8cm]{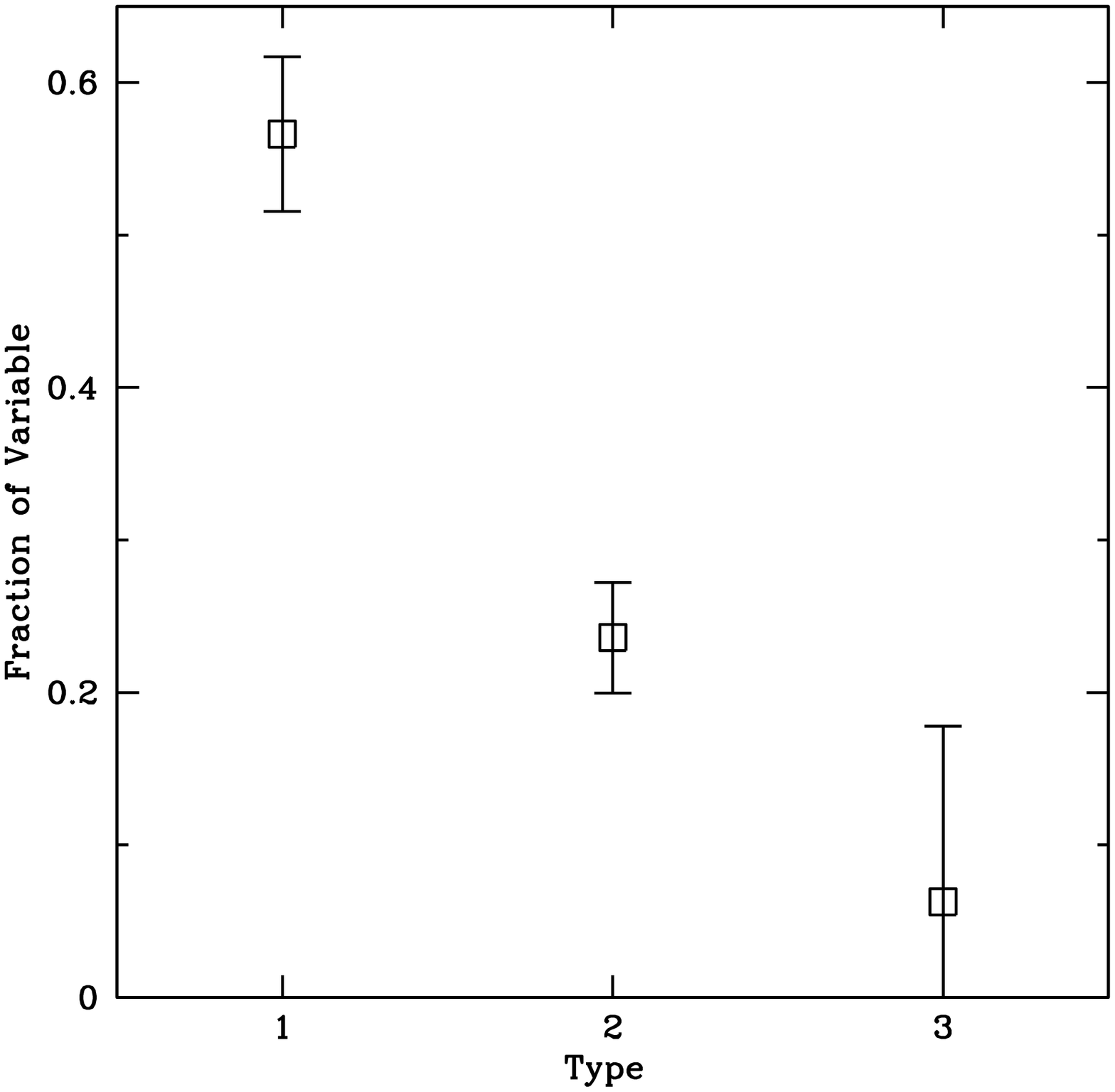}
\caption{{\it Left panel:} Distribution of optical vs X-ray variability for all the sources with $V>1.3$.
{\it Right panel:} Fraction of Optically variable sources, divided by optical type.}
\end{center}
\label{sigmaoptvar}
\end{figure*}

\begin{figure*}
\begin{center}
\includegraphics[width=6cm,height=8cm]{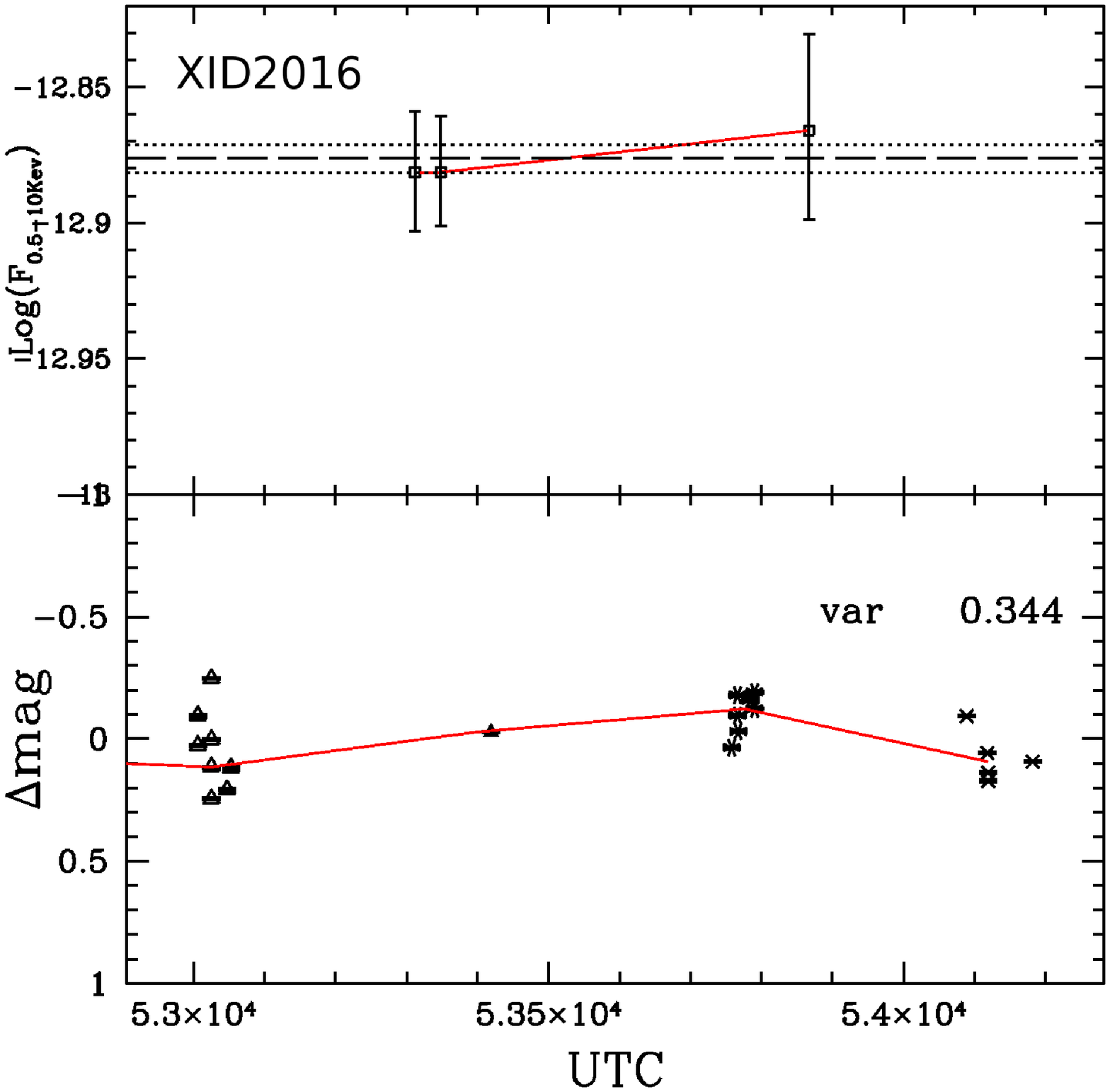}\hspace{1.3cm}\includegraphics[width=6cm,height=8cm]{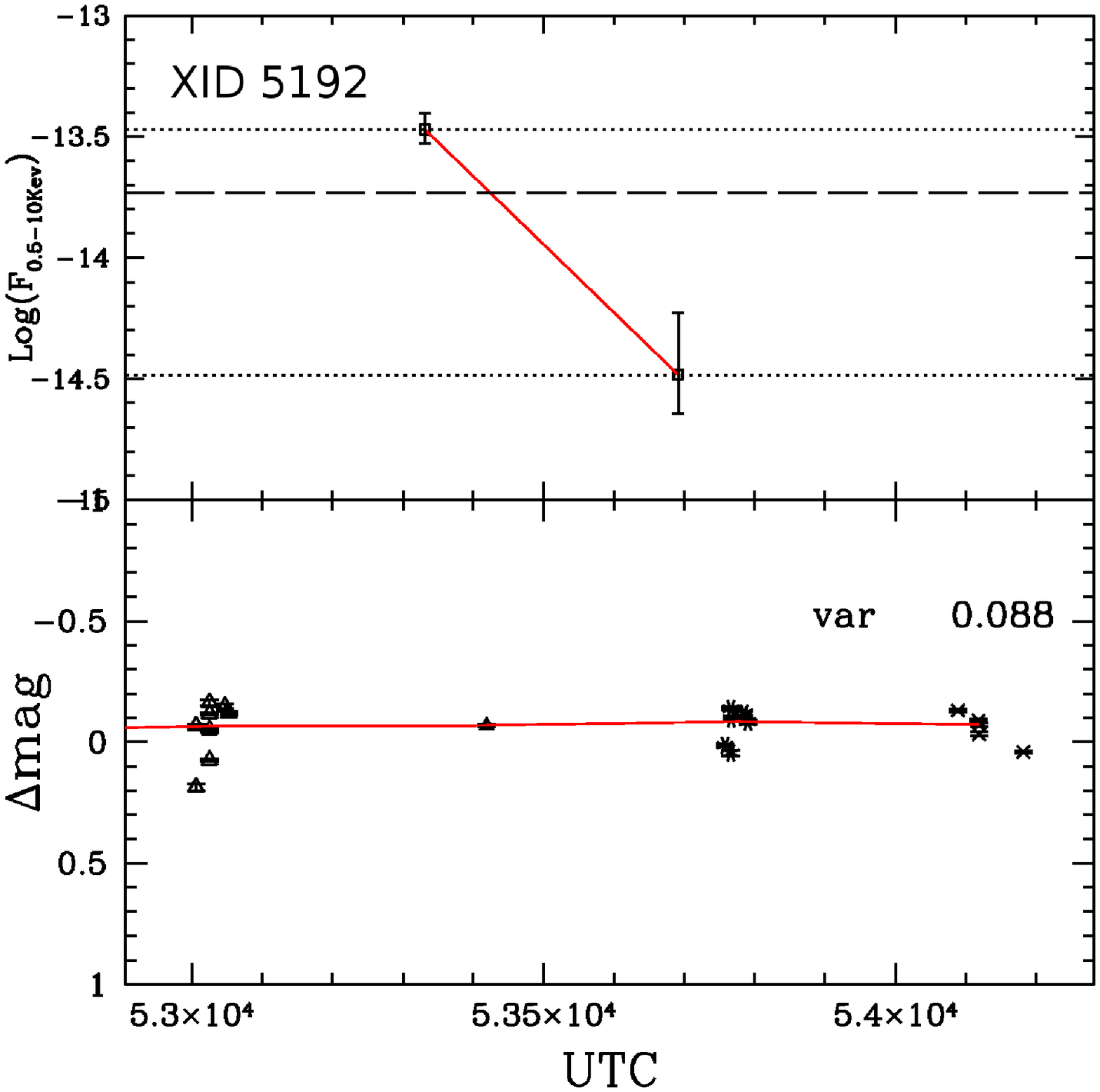}
\caption{{\it Top panels:} \xmm\ 0.5-10 keV, background subtracted light-curve of XID \#2016 and 5192. 
The dashed line shows the weighted mean of the 0.5-10 keV flux. Dotted lines show
the standard error on the mean. The continuous line connects data points.
{\it Bottom panels:} the sources were observed in 4 epochs, each epoch marked with a different symbol, as in Fig. 3. 
The red line connects the median values of the deviation from a running Gaussian filter, for each group of observations. 
The points in each group refer to different wavelengths, therefore the difference within points in each group are due to 
the SED shape, and not variability.
}
\end{center}
\label{lc2}
\end{figure*}

It's also interesting to note that the slope of the \sig\ vs. $M_{\rm BH}$ in Fig. 10 (left)
is the same of the global slope between \sig\ and \lum\ in Fig. 9 (right).
This suggest that the latter is a byproduct of the former one, as observed at lower frequencies (P12).
This is shown if Fig. 11, where the distribution of \sig\ vs. \lum\ is shown, for sources with $>700$ counts,
after normalizing \sig\ for the $M_{\rm BH}$. The linear regression between the \sig\ normalized for the BH mass
is fully consistent with 0 (slope of $0.13\pm0.12$).

The residual scatter of \sig\ after accounting for the $M_{\rm BH}$/\lum\ dependency, it's still of $\sim2$ orders of magnitude,
for individual values. This would imply that the normalization PSD below $\nu_b$ is not the same for all the sources, 
as instead usually assumed.
However, as we pointed out in Sec. 5.1, it has been shown in Allevato et al. (2013), that the bias due to sparse sampling can 
be broadly distributed between 0.1 and 10, meaning that we cannot use the observed scatter of individual \sig\ values, 
as a direct probe for the intrinsic scatter of the PSD normalization.

Regarding the correlation between \sig\ and  Eddington ratio, its existence is debated in the literature
(O'Neill et al. 2005; McHardy et al. 2006; Koerding et al. 2007; P12).
At low frequency Young et al. (2012) found a significant anti-correlation between \sig\ and accretion rate
which they interpreted as an artifact of the \sig-\lum\ anti-correlation.
In our dataset the correlation is globally flat. A hint of a bi modality is present if Fig. 10 (right),
with average \sig\ increasing both for very high and very low accretion rates.
However the quality of the data, and the limited sample included here, do not allow to further investigate this issue.


\subsection{Optical vs. X-ray Variability}

The optical variability, expressed in $\Delta$mag, was computed in Salvato et al. (2009; see details in section 3 of that paper 
for the definition of $\Delta$mag) 
in order to correct for its effect when computing photometric redshifts.
The COSMOS optical photometry has been acquired in five epochs distributed over $\sim6$ years. 
Within each epoch (apart from 2005), several individual filter observations were distributed over less than three months, 
covering the whole optical range.   
This allowed to study and correct AGN time variability over timescales of years, while shorter variability cannot be easily addressed. 
A source with $\Delta Mag>0.25$ was considered variable.

The optical variability is show in Fig. 12 (left) as a function of 
the X-ray \sig, for sources with $V>1.3$.
The two quantities are clearly not correlated ($\rho_S$= -0.113 and $P_S= 0.0611$). 
The optical photometry is not simultaneous to the X-ray observations but rather span a larger time scale. 
Also the cadence is different.
These two factors (in addition to variability been produced at various distances from the BH, 
depending on the wavelength) can explain why X-ray and optical variability are not clearly correlated.

Fig. 13 reports two examples of sources that are highly variable in one band and not in the other.
Source 2016 (left) is a source classified as a Broad line from the optical spectrum,
and it is non-variable in X-ray and highly variable in optical.
Source 5192 (right), classified as type-2 from the SED fitting, instead, is variable in X-ray,
while shows a perfectly flat optical lightcurve.

Finally, the optical variability tends to be highly type-dependent (Fig. 12 right): 
the fraction of type-1 optically variable ($\Delta Mag>$ 0.25 Mag) is considerably higher
($\sim55\%$) than that of type-2 ($\sim25\%$).
We underline however that the variability is one of the criteria that enters the process to identify sources with photometric
redshift as AGN-dominated or galaxy-dominated (see flow chart
in Fig. 6 of Salvato et al. 2011 for details on the template class
assignation), thus this result is partly an induced effect.

\section{Conclusions}

We used the repeated \xmm\ observations in the COSMOS field to study the long term (months-years in rest frame) variability
of a large sample of X-ray detected AGN.
We found that:
\begin{itemize}

\item Variability is prevalent in AGN whenever we have good statistic to measure it 
(up to 75\% of variable sources in the $>1000$ counts regime)

\item There is no significant difference in the distribution of variability between type-1 and type-2 sources

\item The anti-correlation between \sig\ and \lum\ has a rather flat slope ($\alpha=-0.25$)
when the total sample of variable sources ($V>1.3$) is considered. If the sample is divided into redshift bins however, 
a steeper anti-correlation is observed in all bins, with both slope and normalization increasing with redshift 

\item The previous result is affected by two main selection effects: the selection in V,
and the correlation between \sig\ and counts, both producing higher average \sig\ at low \lum\ and high z.
If all and only the sources with good statistics ($>700$ counts) are considered, the evolution in z of the 
\sig-\lum\ relation is no more significant, and the relation at low redshift ($z<0.7$), with slope $-0.40\pm0.06$,
perfectly fits the remaining data points

\item Thanks to the wealth of data available in the COSMOS field, for the first 
time we were able study the correlation between long term X-ray variability
and BH masses and Eddington ratios in a large sample of AGN.
If only sources with more than 700 counts are  considered, 
a strong anti-correlation between $M_{BH}$ and \sig\ 
is observed ($\rho_S$= -0.315 and slope of $-0.42\pm0.11$),
while no correlation is found between L$_{Bol}$/L$_{Edd}$ and \sig

\item The anti-correlation between \sig\ and \lum\ disappears if the \sig\ is normalized for 
the $M_{BH}$, suggesting that the one with luminosity is the byproduct of the more intrinsic 
anti-correlation with $M_{BH}$

\item No clear correlation is found between the optical and the X-ray variability
\end{itemize}

The study of long term variability in large samples of high redshift 
AGN will further benefit in the near future from the ongoing extension of the X-ray coverage 
in the COSMOS field with \chandra\ (The COSMOS Legacy Survey, PI Civano), and from the extra 
3 Msec. of observation in the \chandra-DFS, awarded in Cycle 15 (PI Brandt).



\begin{acknowledgements}

We thank the anonymous referee for the useful comments and suggestions, 
which significantly contributed to improving the quality of the publication.
We thank V. Allevato and T. Dwelly for useful discussions.
GL and GH acknowledge support by the German Deutsche Forschungsgemeinschaft, DFG Leibniz Prize (FKZ HA 1850/28-1).
GP acknowledge support via an EU Marie Curie Intra-European fellowship under contract no. FP-PEOPLE-2012-IEF-331095. 
BT acknowledges support by the Benoziyo Center for Astrophysics.
This research has
made use of the NASA/IPAC Extragalactic Database (NED) which
is operated by the Jet Propulsion Laboratory, California Institute
of Technology, under contract with the National Aeronautics and
Space Administration, and of data obtained from the Chandra
Data Archive and software provided by the Chandra X-ray Center
(CXC). Also based on observations obtained with \xmm\, an ESA science mission
with instruments and contributions directly funded by
ESA Member States and NASA.

\end{acknowledgements}

\end{document}